\documentclass[useAMS,usenatbib]{mn2e}

\usepackage{graphicx}

\title[Light curves of WZ Sge stars]{On the nature of quiescent light curves demonstrated by WZ Sge stars}

\author[D. A. Kononov et al]{D. A. Kononov$^{1}$,\thanks{E-mail:\ dkononov@inasan.ru}
\LARGE \rm C. H. S. Lacy$^{2}$, V. B. Puzin$^{1}$ V. P. Kozhevnikov$^{3}$,\\
\ \\
\LARGE \rm A. Yu. Sytov$^{1}$, A. P. Lyaptsev$^{3}$\\
$^{1}$Institute of Astronomy of RAS, 48 Pyatnitskaya Str., Moscow, 119017, Russia\\
$^{2}$University of Arkansas, Fayetteville, AR 72701, USA\\
$^{3}$Astronomical Observatory, Ural Federal University, 51 Lenin Ave., Ekaterinburg, 620083, Russia}
 
\begin{document}

%\date{Accepted. Received; in original form}

\pagerange{\pageref{firstpage}--\pageref{lastpage}} \pubyear{}

\maketitle

\label{firstpage}

\begin{abstract} 
We present the results of simultaneous spectroscopic and photometric observations of the cataclysmic variable star (hereafter CVs) V455 Andromedae, belonging to the WZ Sge sub-class, in quiescence. Using the spectroscopic data we computed time-resolved Doppler tomograms of the system demonstrating its behavior at different orbital phases. In the tomograms one can see the periodic brightening of different regions within one orbital cycle. We interpret this brightening as being due to the interaction of four phase-locked shock waves in the disk with a specific internal precessing density wave that develops inside the disk, because of the tidal influence of the secondary star, and this density wave propagates up to the disk's outer regions. The shock waves, located in the outer regions of the disk, are: two arms of the tidal shock; the "hot line", a shock occurring in the region of the interaction between the stream from the $L_{1}$ point and the disk; and a bow-shock, occurring ahead of the disk due to its orbital supersonic motion in the circumbinary gas. When the outer part of the density wave in its precessional motion reaches a shock wave the local density grows, which amplifies the shock (by increasing $\rho V^{2}/2$). This results in an additional energy release in the shock and can be observed as a brightening. Analysis of the tomographic results and the photometric data shows that two main sources contribute to the light curves of the system: the radiation of the "hot line" and the bow-shock gives us two major orbital humps, located approximately at the orbital phases $\phi=0.25$ and $\phi=0.75$; the amplification of the four shock waves may give us up to four "superhumps" shifting over the light curve with the precessional period. These two effects, when overlapping, change the shape of the light curves, shift the hump maxima, and they sometimes produce more than two humps in the light curve. We should emphasize that when saying "superhumps" we imply an effect that is observed in quiescent light curves of WZ Sge stars, as opposed to ''classical'' superhumps usually observed in outbursts.
\end{abstract} 


\section{INTRODUCTION}

WZ Sge stars (sub-class of SU UMa stars) are evolved close binary systems comprised of late type secondary components, filling their Roche lobes, and white dwarfs as primary components. To date, there have been about 60-100 representatives of this class discovered (see, e.g., \citet{katy2014, goly2014}). Specific orbital periods of these objects are near the period minimum (usually, $\approx$80 min.), and mass ratios are extremely low ($q=M_{2}/M_{1}<0.1$). In some works (see, e.g., \citet{araj} and references inside) the authors suppose that some representatives of this sub-class can even harbor brown dwarfs, no longer undergoing fusion reactions, as the secondary. One of the most noticeable observational features of the WZ Sge stars is their very powerful and rare (recurrent periods up to decades) super-outbursts when a star increases its brightness by over $6^{m}$. "Normal" outbursts in these systems are very rare or even absent (\citet{hara}). The main photometric observational features of these objects on short time scales (one orbital period) are the so-called double-humped light curves (see, e.g., \citet{araj, katy2009}), demonstrated in quiescence, i.e. one observes two pronounced humps within a single orbital period of a WZ Sge type system.

There is still no commonly accepted model about how WZ Sge orbital light curves appear. Some authors  (see, e.g., \citet{avil}, \citet{tovm}) suppose that the two humps may be caused by radiation from two arms of a tidal shock or density waves, formed in the disk under the action of tidal resonances. The shape and position of the humps in this model are determined by the viewing aspect of each arm. Another model proposes the ricochet of the gas stream, issuing from the inner Lagrangian point, and the formation of two bright regions in the disk, one "classical" hot spot and another on the opposite side of the disk (see \citet{wolf, Silv12}). In the frame of this model two bright spots give two humps in the light curve.

However the proposed models cannot explain a number of effects, observed in WZ Sge stars. One of these effects is the varying number of the humps observed within one orbital period. For example, \citet{kitsi} report the results of long-term monitoring of the WZ Sge system, the sub-class prototype, and mention that in different observational epochs they see from two to three pronounced humps. Studying the system SDSS J080434.20+510349.2, related to WZ Sge stars, \citet{pavl2009} reports four (!) humps. Sometimes the reported number of the humps may be reduced to only one per orbital period (\citet{araj, katy2009}). This also means that it would be more correct to call the light curves of WZ Sge stars multi-humped instead of double-humped. In addition, the humps may shift over binary phases from one observational epoch to another and change their shape.

The observational facts, mentioned above, contradict the proposed models of two tidal arms or two bright spots. First of all, no one of them can explain the varying number of the humps. Then, the results of the numerical simulations (see, e.g. \citet{smh_1, smis, BiBo04, Bi05, KuBi01}), conducted using grid-based methods, and the results of Doppler tomography (see e.g. \citet{shh97}) predict that the arms of the tidal shock in the disks of CVs are phase-locked. Indeed, if these spiral shocks had not been phase-locked we wouldn't have seen them in Doppler tomograms as separate bright regions, since, by definition, any flow element moving in the reference frame, co-rotating with the binary, should be smeared into a ring in the resulting tomogram. Therefore, if we suppose that the humps occur due to the viewing aspects of these features, we should always see them at the same phases.

Recently (\citet{ko2015}) proposed a model, based on the results of 3D gas dynamic simulations of the V455 Andromedae system, that can explain almost all the properties of double-humped light curves including the varying number of the humps and their shift over orbital phases. Within this model we interpret the observed effects by the interaction of four phase-locked shock waves in the disk with a specific internal precessing density wave that develops inside the disk due to the tidal influence of the secondary star, and propagates up to its outer regions. We discuss the model in detail below.

In this work we aim to find observational evidence that our model is correct. The structure of the paper is as follows. In Section 2 we describe our gas-dynamical model. In Section 3 we describe our spectroscopic and photometric observations. Section 4 is focused on analysis of observational Doppler tomograms, calculated using the obtained profiles of the $H_{\alpha}$ emission line. In Section 5 we compare the observational and synthetic Doppler tomograms and analyze the photometric data. In the section "Conclusions" we summarize the results of our study.

\section{GAS-DYNAMICAL MODEL}

\begin{figure}
\includegraphics[width=84mm]{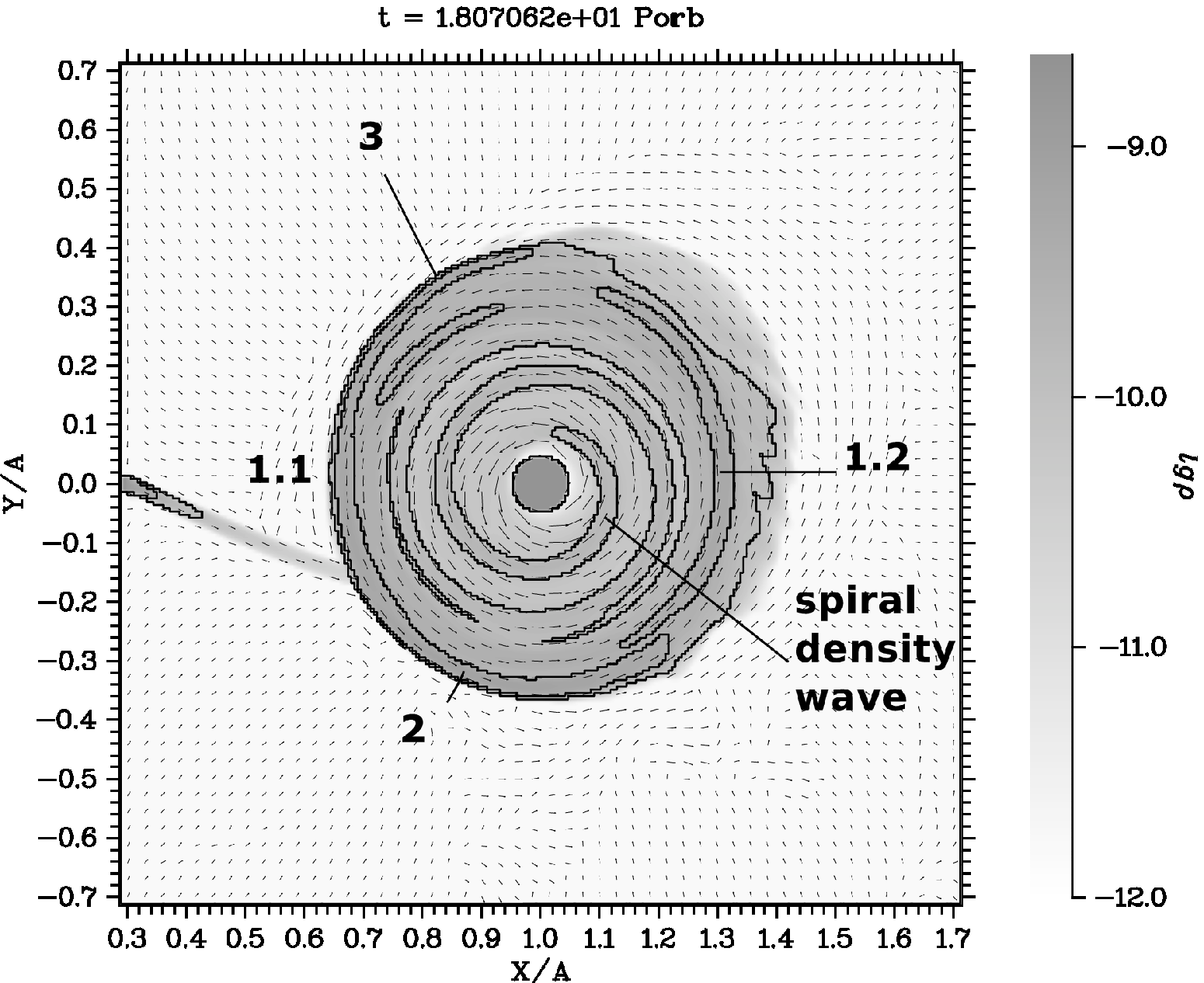}
\includegraphics[width=84mm]{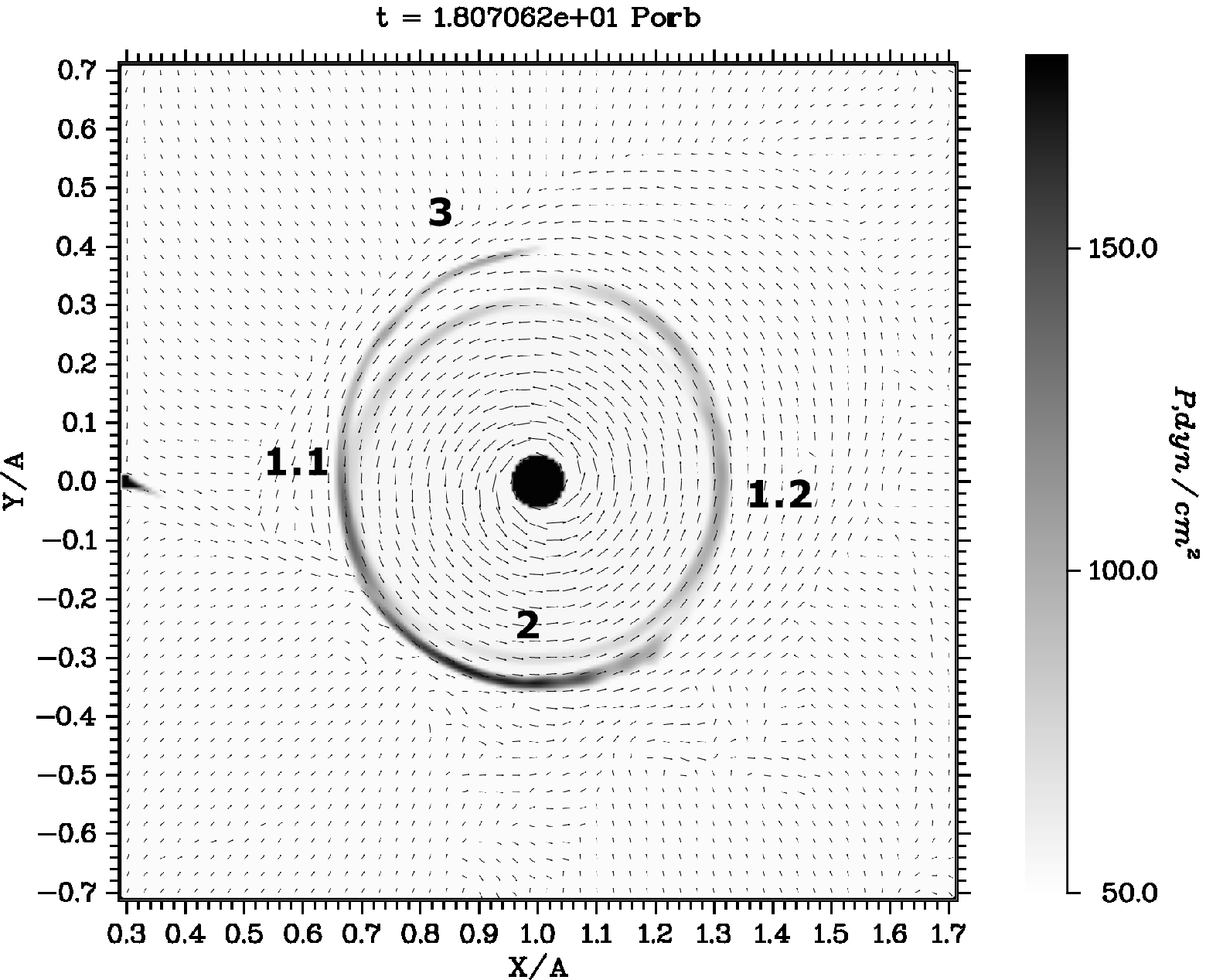}
\caption{Modeled density (right-hand panel) and pressure (left-hand panel) distributions in the accretion disk of V455~And. The axes units are given as a fraction of the binary separation $A=0.56 R_{\odot}$ (\citet{araj}). The inner Lagrangian point in both the panels is on the left. Four major shock waves in the disk are designated by numbers: 1.1 and 1.2 are the arms of the tidal shock; 2 is the hot line; 3 is the bow-shock.}
\label{Fig1}
\end{figure}

To investigate the flow structure in the V455 And system we conducted 3D gas-dynamical simulations using a grid-based method. The method and model are described in detail by \citet{ko2015}. Here we briefly introduce the concept of the model. Our simulations show that a planar accretion disk forms in the system (Fig.~\ref{Fig1}). In the disk four major shock waves occur. These are: two arms of the tidal shock wave (designated by 1.1 and 1.2 in Fig.~\ref{Fig1}); the hot line\footnote{In the classical observational works this shock wave is referred as the hot spot, but hereafter we prefer to call it the "hot line", since the physical mechanism of its formation and its shape (see, e.g. \citet{KuBi01, BiBo04, Bi05}) differ from those initially proposed for the hot spot.} (designated by 2 in Fig.~\ref{Fig1}), a shock wave occurring because of the interaction of the circum-disk halo (outer disk) and the gas stream from the inner Lagrangian point; and the bow-shock (designated by 3 in Fig.~\ref{Fig1}) caused by the super-sonic motion of the accretor and disk in the gas of the circum-binary envelope. Besides, we can observe a one-armed spiral density wave (see Fig.~\ref{Fig1}), starting in the vicinity of the accretor and propagating to the outer regions of the disk. According to \citet{BiBo04, Bi05} this wave occurs in a planar accretion disk due to the retrograde apsidal motion of streamlines.

We should note that density waves were reported in early papers on SPH gas-dynamical simulations (see e.g. \citet{white88, simp98}). However, results obtained with grid-based methods appeared drastically different (\citet{steh99, smh_1, mama98}). For instance, \citet{steh99} reports shock waves instead of density waves in the outer regions of the disk and discusses possible issues of the SPH methods that may give spurious results. Thus, hereafter, we base our considerations only on the simulation results obtained with the 3D grid-based gas-dynamical methods that demonstrate shock waves in the outer regions of the disk and the density wave in its inner regions.

\begin{figure*}
\includegraphics[width=84mm]{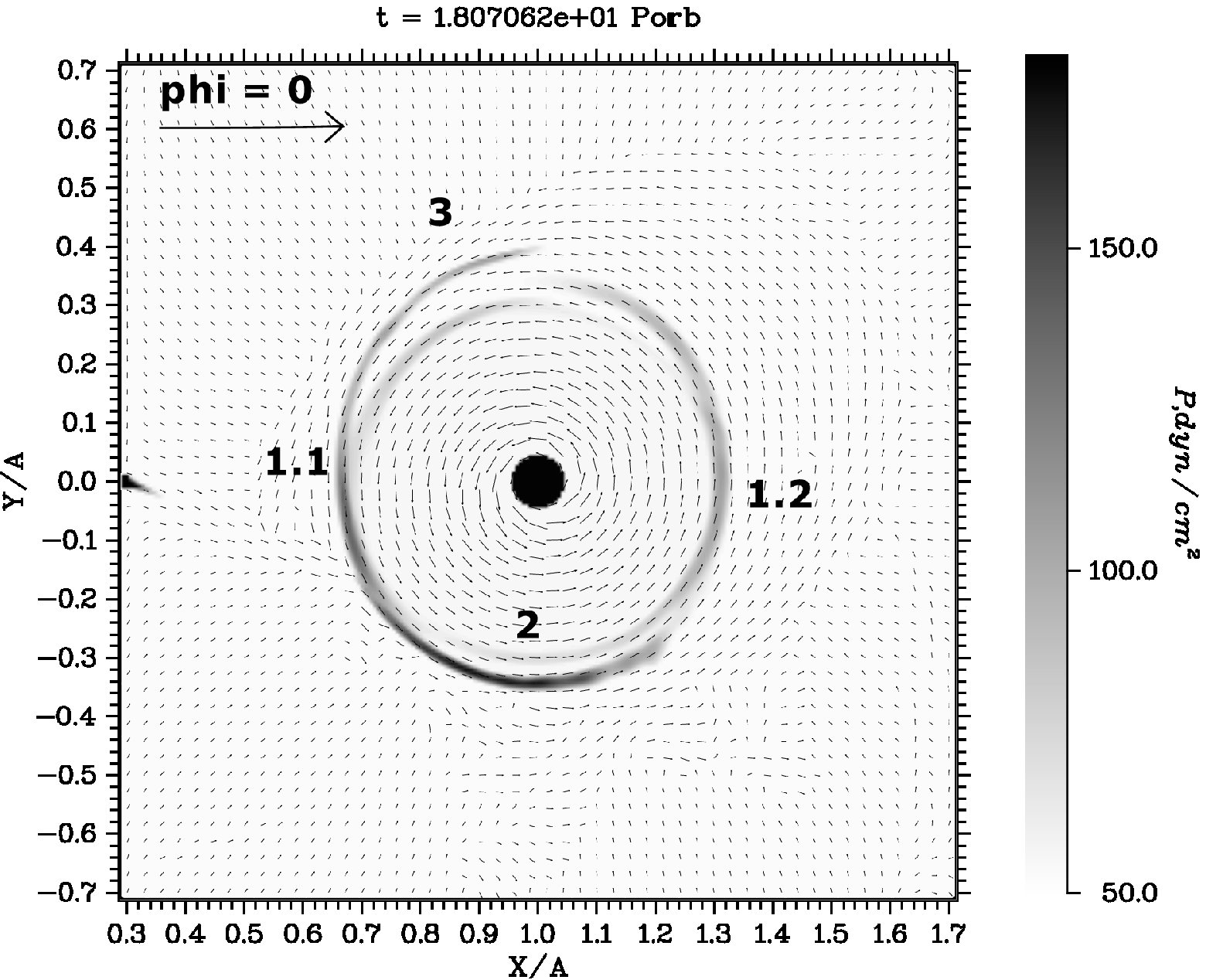}
\includegraphics[width=84mm]{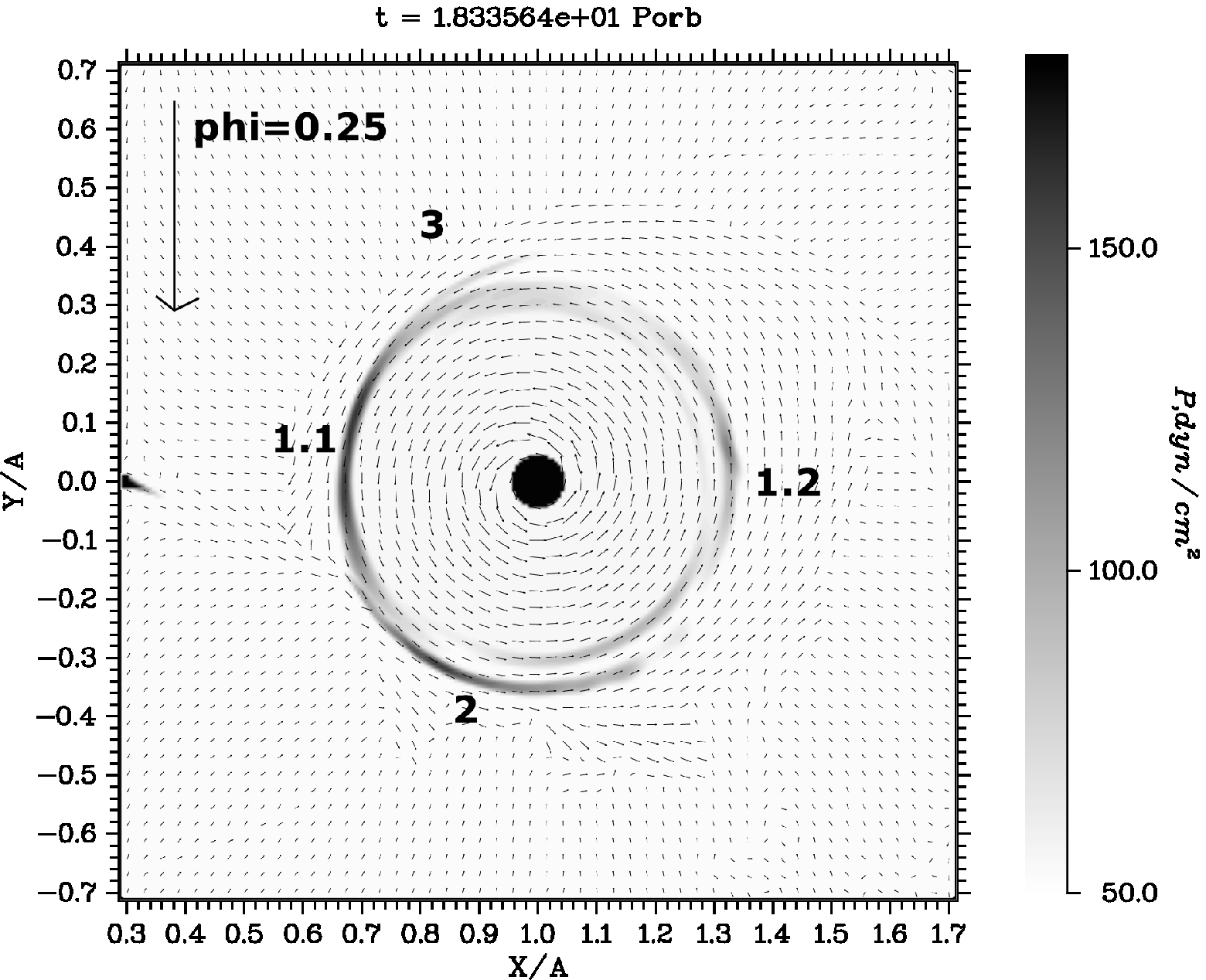}\\
\includegraphics[width=84mm]{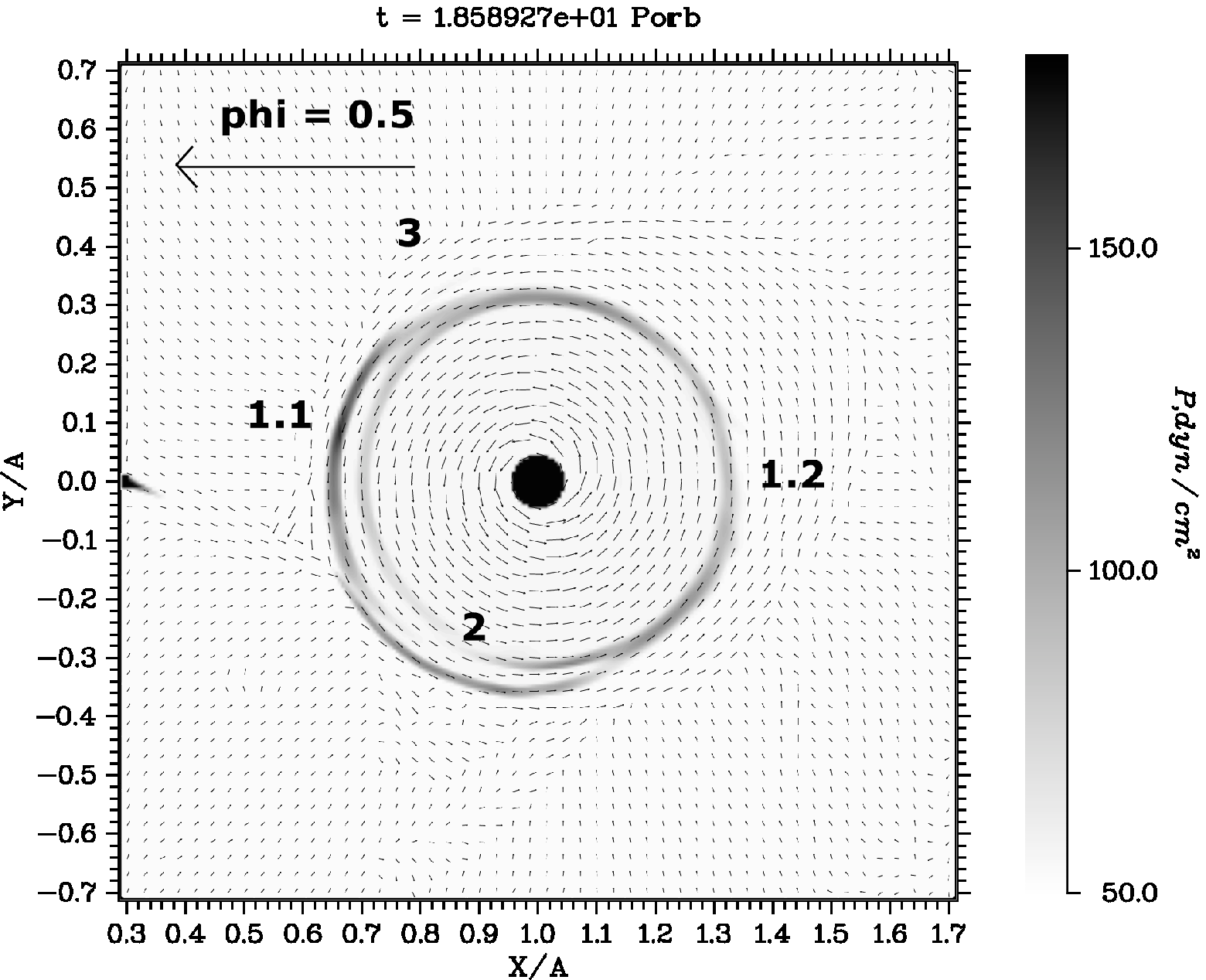}
\includegraphics[width=84mm]{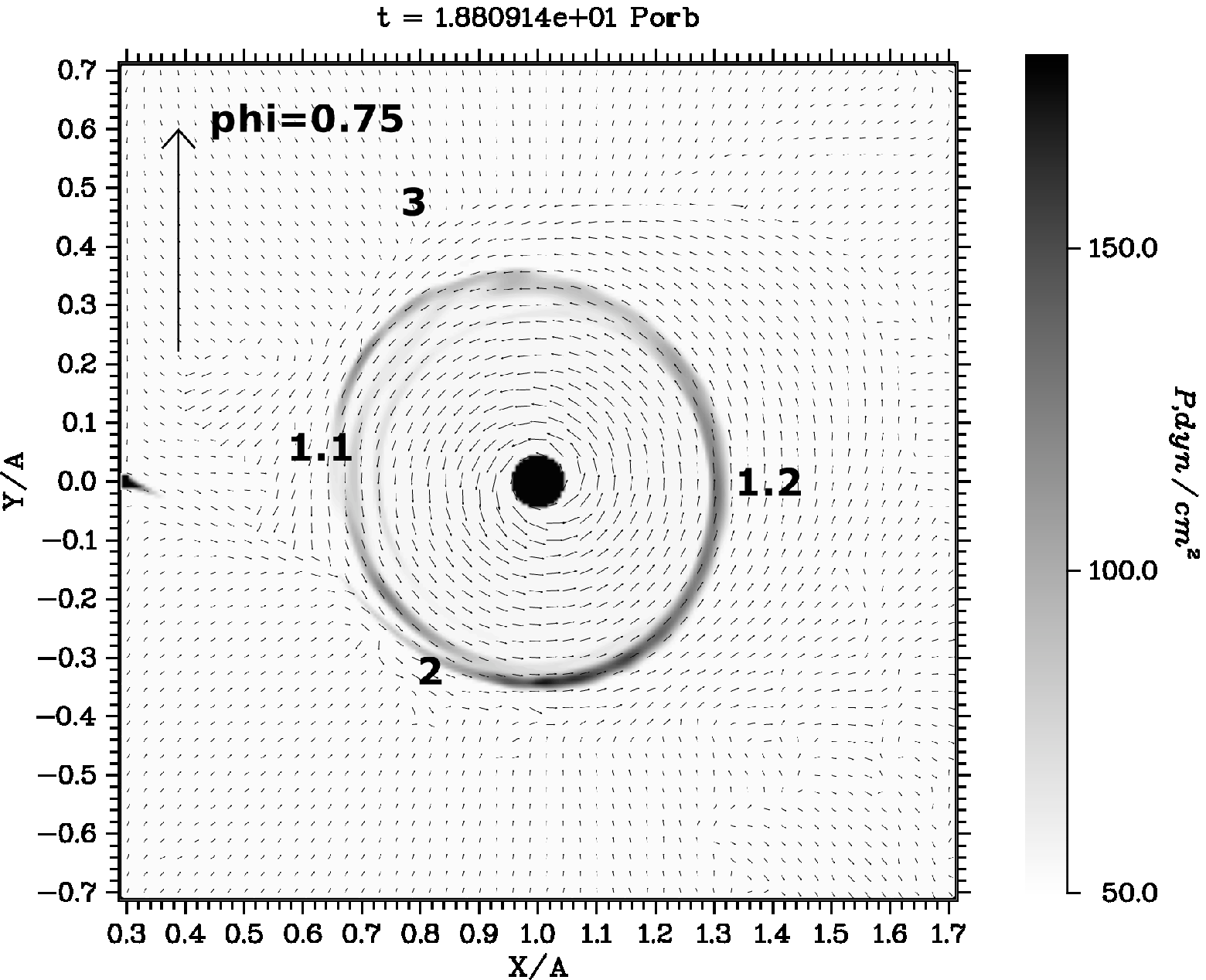}
\caption{Modeled pressure distributions in the accretion disk of V455 And for four subsequent time moments $\tau = 18.07, 18.34, 18.59,$ and $18.81~P_{orb}$. The shock waves are designated in the same way as in Fig.~\ref{Fig1}. Arrows in the panels demonstrate the line of sight at the given moment. Corresponding binary phases are written.}
\label{Fig2}
\end{figure*}

Under the tidal action of the secondary component the density wave retrogradely precesses with an angular velocity that differs from the orbital angular velocity of the system by only 1-2\%. This means that in the observer's (or inertial) coordinate frame this density wave almost rests while the binary system rotates. Here we need to note that we do not speak about a precession of a tilted disk. We speak about the retrograde apsidal motion of a structure, including the planar disk and its inner spiral density wave, and call this process "precession" just for the sake of shortness. This is the crucial point of our model. We should also emphasize that in the simulated accretion disk of V455~And the precessional density wave propagates up to the outer regions of the disk. In the course of the disk's orbital rotation each of the four shock waves passes through the outer part of the density wave. This results in an increase in the  periodic local density in the region of each shock wave. Since the energy, released in shock waves as radiation, depends on the kinetic energy of gas $\rho V^{2}/2$, by increasing the density $\rho$ this effect increases the energy release in the shock wave, which may be observed as an increased brightness or a hump in the light curve. The effect is clearly visible in Fig.\ref{Fig2} where we show four simulated pressure distributions separated by $\approx0.25 P_{orb}$. In the panels of Fig.\ref{Fig2} one can see the consequent amplification of each shock wave within one orbital period. In the top-left panel we see the amplification of the hot line (2). Then in $0.25 P_{orb}$ one of the arms of the tidal shock (top-right panel, 1.1) is amplified. Further the pressure subsequently grows in the regions near the bow-shock (bottom-left panel, 3) and the other arm of the tidal shock (bottom-right panel, 1.2). This effect gives us four potential humps in the light curve. The number of the humps may be reduced to two (which is often observed) because the contribution of some waves may overlap or some shock waves may be initially stronger than the other and their contribution prevails. The shift of the humps over orbital phases in the frame of our model can be explained by the different angular velocities of the precession and orbital motion. Since the precessional angular velocity differs from the orbital angular velocity, in the next orbital cycle a shock wave approaches the outer part of the density wave at a slightly different moment, i.e. we should see the hump slightly shifted over the orbital phase. Within one night of observations this effect may be negligible, since the difference in the velocities is very small but it may be significant, as we show below, when a system is observed in different epochs.

We should note that within our model the retrograde precession should cause negative quiescent superhumps in the light curves of WZ Sge-type stars. Some authors (see e.g. \citet{white88, simp98}) in early works reported progradely precessing SPH-simulated disks. However, later grid-based simulations of planar disks (see \citet{steh99, BiBo04, Bi05}) have never shown prograde precession.

\section{OBSERVATIONS AND DATA REDUCTION} 
\subsection{Photometry}

\rm In our photometric observations of V455~And, we used a multi-channel pulse-counting photometer with photomultipliers that allows us to make continuous brightness measurements of two stars and the sky background. Since the angular separation between the program and comparison stars is small, such differential photometry allows us to obtain magnitudes, which are corrected for first order atmospheric extinction and for other unfavorable atmospheric effects (unstable atmospheric transparency, light absorption by thin clouds, etc.). Moreover, we use a CCD guiding system, which enables precise centering of the two stars in the diaphragms to be maintained automatically. This greatly facilitates the acquisition of long continuous light curves and improves the accuracy of brightness measurements. The design of the photometer is described by \citet{kozhevnikoviz}.

V455~And was observed photometrically in September 2014 during 5 nights by using the 70-cm telescope at the Kourovka Observatory of the Ural Federal University (Russia). The journal of the observations is given in Table~\ref{journal}. The program and comparison stars were observed through 16-arcsec diaphragms, and the sky background was observed through a 30-arcsec diaphragm. The comparison star is USNO-A2.0 1275-18482440. It has $\alpha=23^h33^m40\fs79$, $\delta=+39\degr17\arcmin24\farcs6$ and $B=14.7$~mag. Data were collected at 1-s sampling intervals in white light (approximately 300--800~nm), employing a PC-based data-acquisition system.

\begin{table}
{\small 
\caption{Journal of the photometric observations.}
\label{journal}
\begin{tabular}{@{}l c c}
\hline
\noalign{\smallskip}
Date  &  HJD start & length \\
(UT) & (-2466900) & (h) \\
\hline
2014 September 22   & 23.195645 &  7.0  \\
2014 September 23  & 24.194649 & 3.0  \\
2014 September 24   & 25.166714 & 8.1   \\
2014 September 25  & 26.172374 & 7.2  \\
2014 September 27   & 28.160955 & 7.2  \\
\hline
\end{tabular} }
\end{table}

\begin{figure}
\includegraphics[width=84mm]{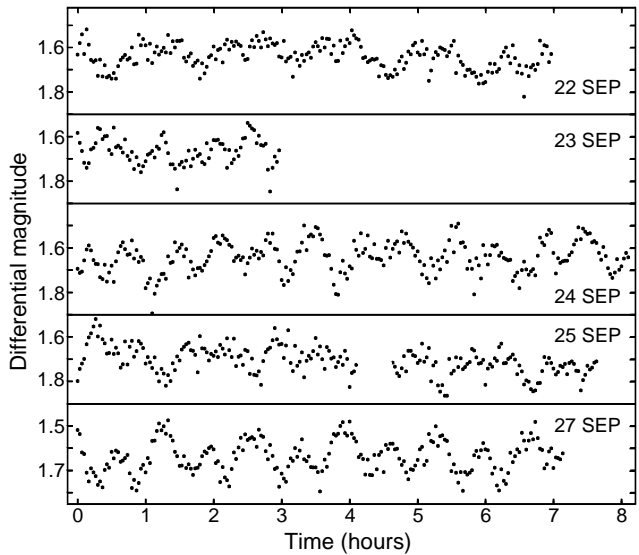}
\caption{Differential light curves of V455~And for five nights of observations from September 22 to September 27, 2014 except the night of September 26 when the weather conditions were poor.}
\label{Fig3}
\end{figure}

We obtained the differences of magnitudes of the program and comparison stars taking into account the differences in light sensitivity between the various channels. According to the mean counts, the photon noise (rms) of the differential light curves is equal to 0.21--0.26~mag (a time resolution of 1~s). The actual rms noise also includes atmospheric scintillations and the motion of the star images in the diaphragms. But these noise components contribute insignificantly. Fig.~\ref{Fig3} presents the differential light curves of V455~And, with magnitudes averaged over 120-s time intervals. The white noise component of these light curves is 0.019--0.024~mag.

Using a fast Fourier transform we analyzed the periodicities in the light curves. The main periodicity found is $P = 81.080\pm0.020$ min., which corresponds to the orbital period of the system previously reported by \citet{araj}. Besides, we have a well pronounced period $P_(sh) = 80.391\pm0.07 $ min. that can be attributed to the superhumps. The errors in the periods were calculated using the method of Schwarzenberg-Czerny (\cite{schw91}).

We should note that the obtained superhump period is shorter than the orbital period, which is contrary to the results of \citet{araj} where the superhump period was $P_{AB} = 83.38$ min. Besides, \citet{kozh15} argues that the superhump period reported by \citet{araj} may be spurious due to the effect of extremely large gaps between their photometric observations. Though this issue requires additional analysis, hereafter we discuss only negative quiescent superhumps of V455 And, since, as mentioned above, the results of gas-dynamical simulations also demonstrate only the retrograde precession that must result in negative superhumps.

\subsection{Spectroscopy}

\begin{figure}
\includegraphics[width=84mm]{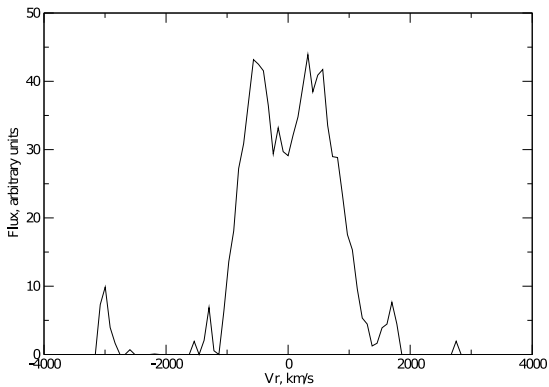}
\caption{An $H_{\alpha}$ profile from 60 profiles, obtained on September 22, 2014}
\label{Fig4}
\end{figure}

Along with the photometric observations we conducted simultaneous spectroscopic observations at the Terskol Observatory (Terskol branch of The Institute of Astronomy of the Russian Academy of Sciences) using the Multi-Mode Cassegrain Spectrograph of the 2-meter Zeiss-2000 telescope. We attempted to observe the system within the same 5 nights in September 2014. However, unfortunately, only one observational night of September 22 was successful because of the weather conditions. We conducted our observations in the classical mode with the spectral resolution $R\approx2000$. The covered wavelength range containing the $H_{\alpha}$ line was $6000 - 7000 \AA$. The exposure time per one spectrum was 6 minutes. In total, on September 22, 2014 we obtained 60 individual spectra, covering over 5 orbital periods of V455 And (10 - 12 spectra per orbital period). The average S/N ratio of the obtained spectra is 15, which is not much, since the system is quite faint ($\approx 16^{m}$) and the exposure times are relatively short.

For data reduction we used the standard package ESO MIDAS (context Long). The reduction procedure included bias subtraction, flat-fielding and the sky background subtraction. The data were continuum subtracted using a constant only, since the continuum was flat under the $H_{\alpha}$ line. At the final step we performed the wavelength calibration of each individual spectrum and recalculated wavelengths to radial velocities. As a result we obtained 60 profiles of the $H_{\alpha}$ line. One of the obtained profiles is shown in Fig.~\ref{Fig4}.

\section{DOPPLER TOMOGRAPHY}

\rm To analyze the obtained spectroscopic data we used the technique of Doppler tomography (\citet{Marsh88}). Within this technique emission line profiles, observed at subsequent orbital phases of a cataclysmic binary star, are considered as tomographic projections of the intensity distribution over velocity space. By restoring a Doppler tomogram we obtain the image of the accretion disk in velocity coordinates. The tomogram is more illustrative and easier for analysis than the initial spectra. In the tomogram we can see the main gas-dynamical features as, for example, shock waves in the accretion disk. 

Since we want to examine our model, predicting the periodic brightening of the shock wave regions in the accretion disk of V455 And, we decided to calculate the Doppler tomograms by using line profiles covering 10 subsequent ranges of the system's orbital cycle, namely $\Delta \phi \approx 0.0 - 0.5; 0.1 - 0.6; 0.2 - 0.7; ... 0.9 - 1.4$. If our model is correct we should obtain 10 different Doppler tomograms, since each shock wave brightening must be visible within a certain limited time interval.

To calculate the tomograms we phase-binned our 60 line profiles, initially covering the range of 5 orbital periods, as though they had been obtained within the phase interval $\Delta \phi \approx 0.0 - 1.4$. This means that, for example, a profile, obtained at the phase of $\phi = 1.6$ is associated with the phase $\phi = 0.6$ in the new set or  $\phi = 2.8$ becomes $\phi = 0.8$. To phase-bin the profiles we used the ephemeris calculated by \cite{araj}:

$$HJD 2451812.67765(35)+0.05630921(1)\times E$$

\begin{figure*}
\includegraphics[width=84mm]{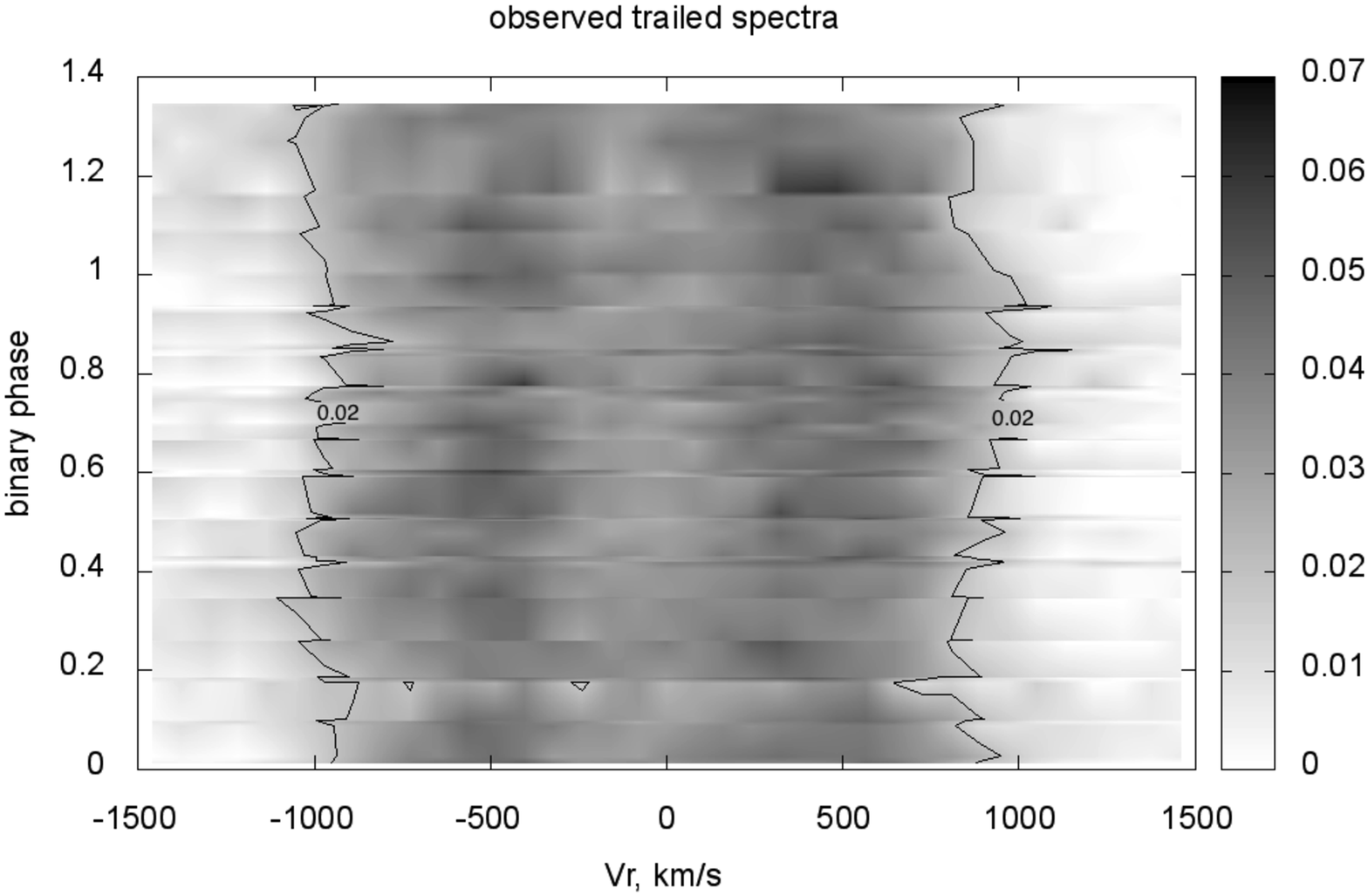}
\includegraphics[width=84mm]{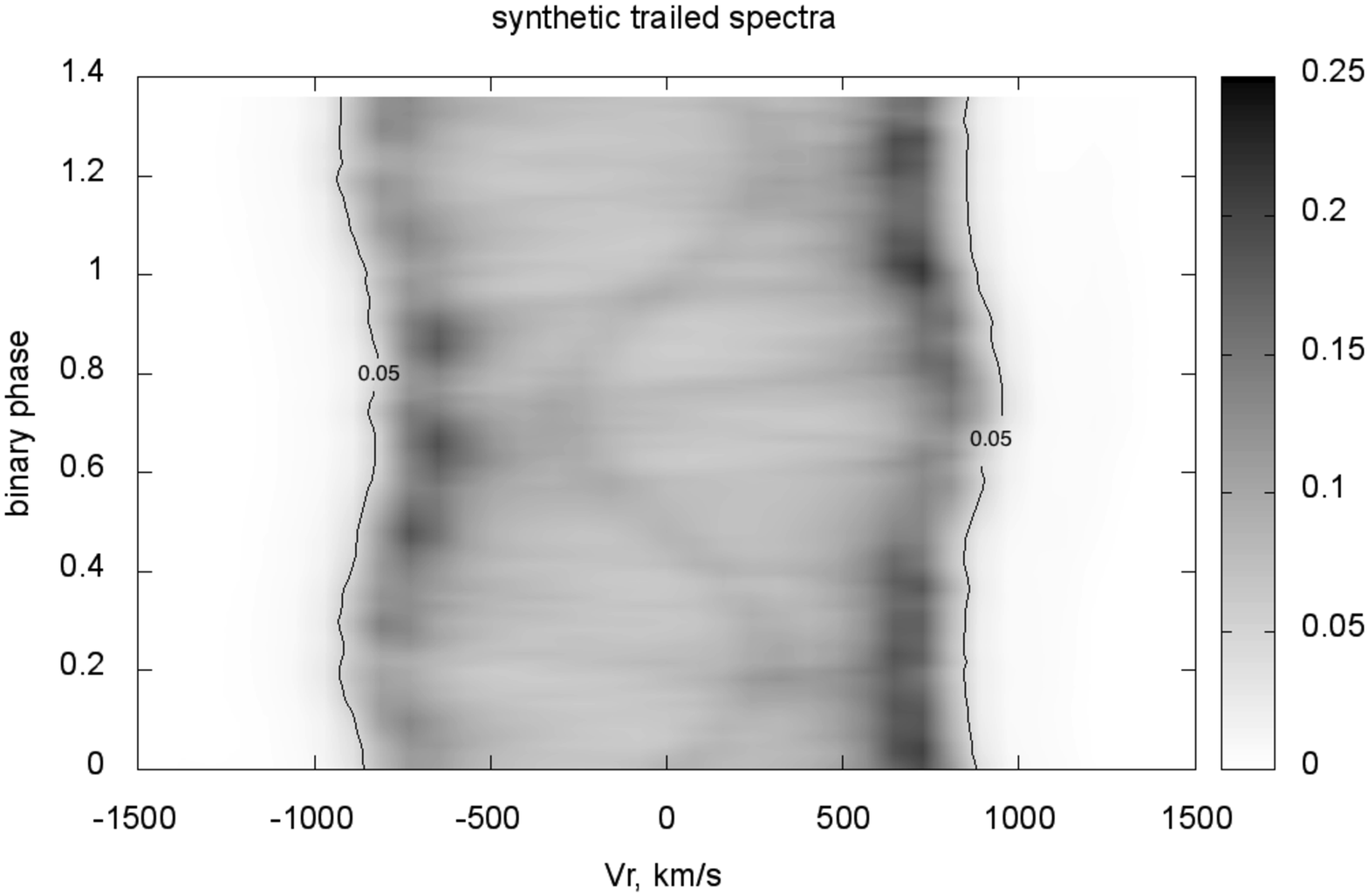}
\caption{Observed (left-hand panel) and synthetic (right-hand panel ) trailed spectra of V455 And. The observed trailed spectra include all the 60 profiles.}
\label{Fig5}
\end{figure*}

We should note that by collapsing the initial sample of profiles, obtained within the phase interval of 5 orbital periods, into the interval $\Delta \phi \approx 0.0 - 1.4$ we may bring a certain error in the final tomograms, since the disc structure changes from one orbital cycle to the next due to the precession. However, according to the following expression $P_{am} = \frac{P_{orb}\times P_{sh}}{P_{orb} - P_{sh}}$ (see, e.g., \cite{warn}), the precessional period of V455 And is $P_{am} \approx 6.57$ days or $\approx 116 P_{orb}$. This allows us to suppose that in 5 orbital periods the disk structure doesn't change too much. The resulting trailed spectra are shown in the right-hand panel of Fig.~\ref{Fig5}

After re-binning we divided the 60 line profiles in 10 sub-sets as mentioned above. This means that an individual profile may be included in several sets, depending on its orbital phase. For example a profile associated with $\phi=0.2$ is included in the sets of $\Delta \phi \approx 0.0 - 0.5; 0.1 - 0.6; 0.2 - 0.7$. Finally, each set contained $\approx25$ line profiles, which allowed us to increase the S/N ratio of final tomograms. To calculate the tomograms we used our own maximum entropy code, based on the regularized algorithm proposed by \cite{Lucy94}.

\begin{figure*}
\includegraphics[width=55mm]{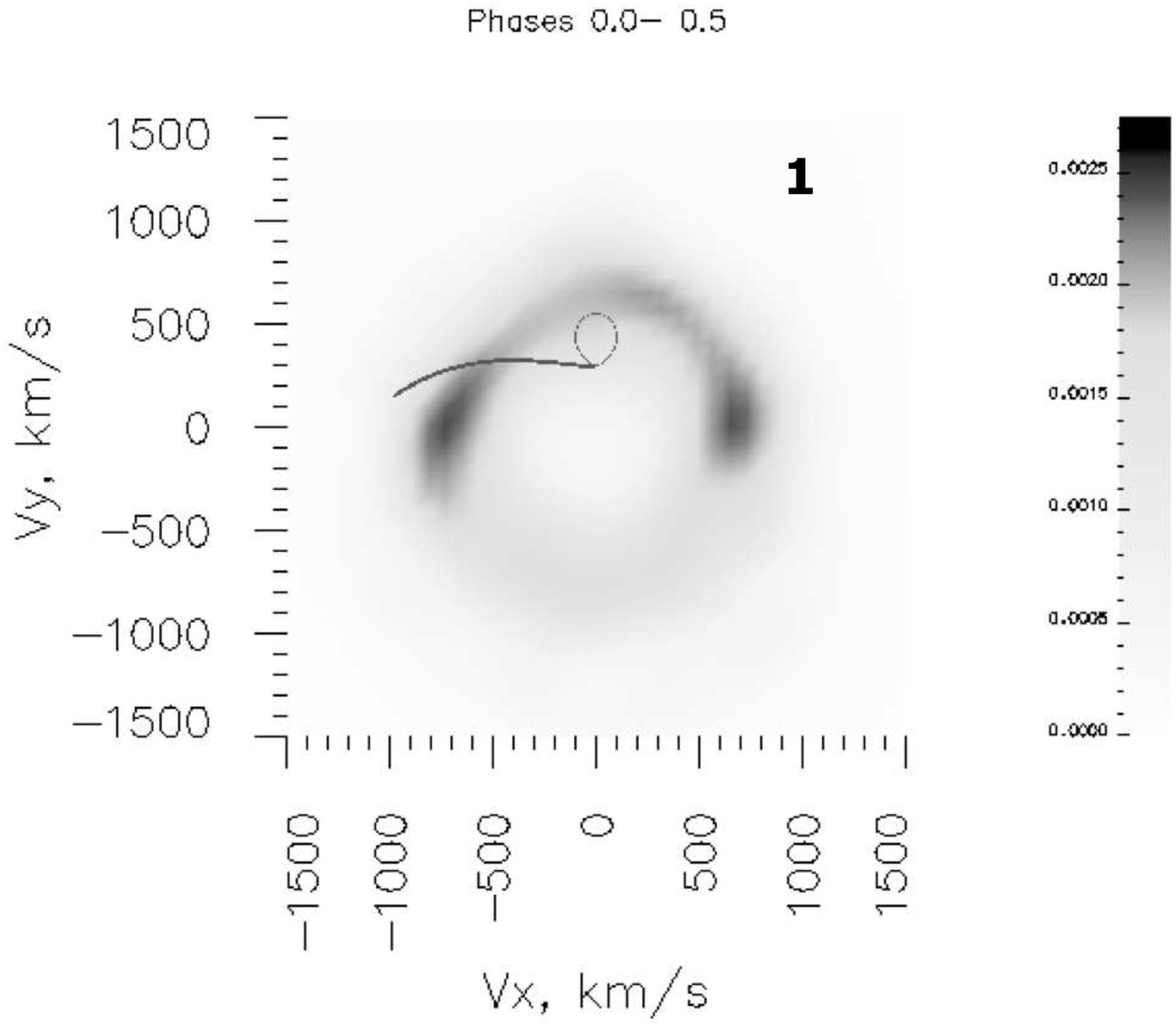}
\includegraphics[width=55mm]{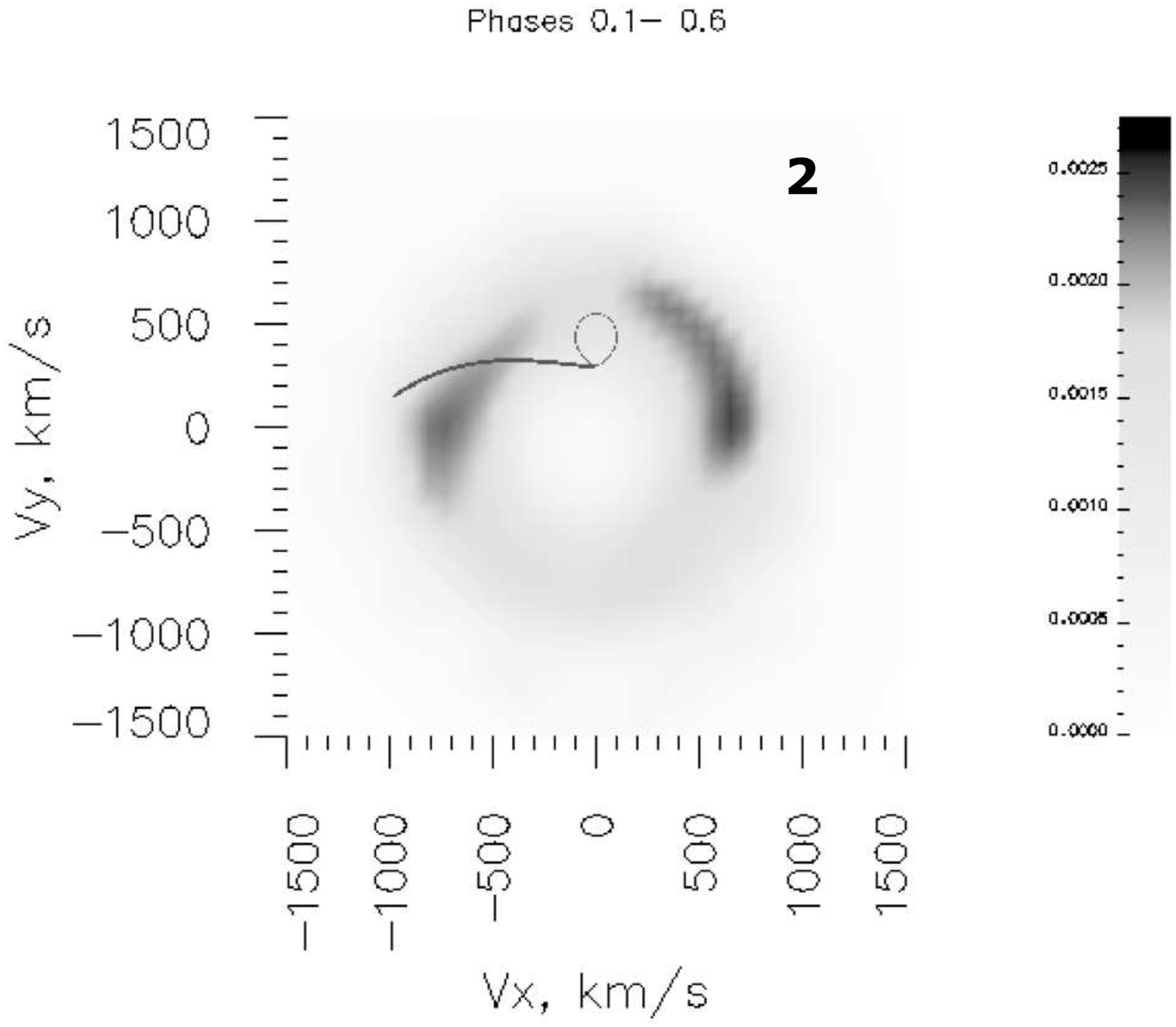}
\includegraphics[width=55mm]{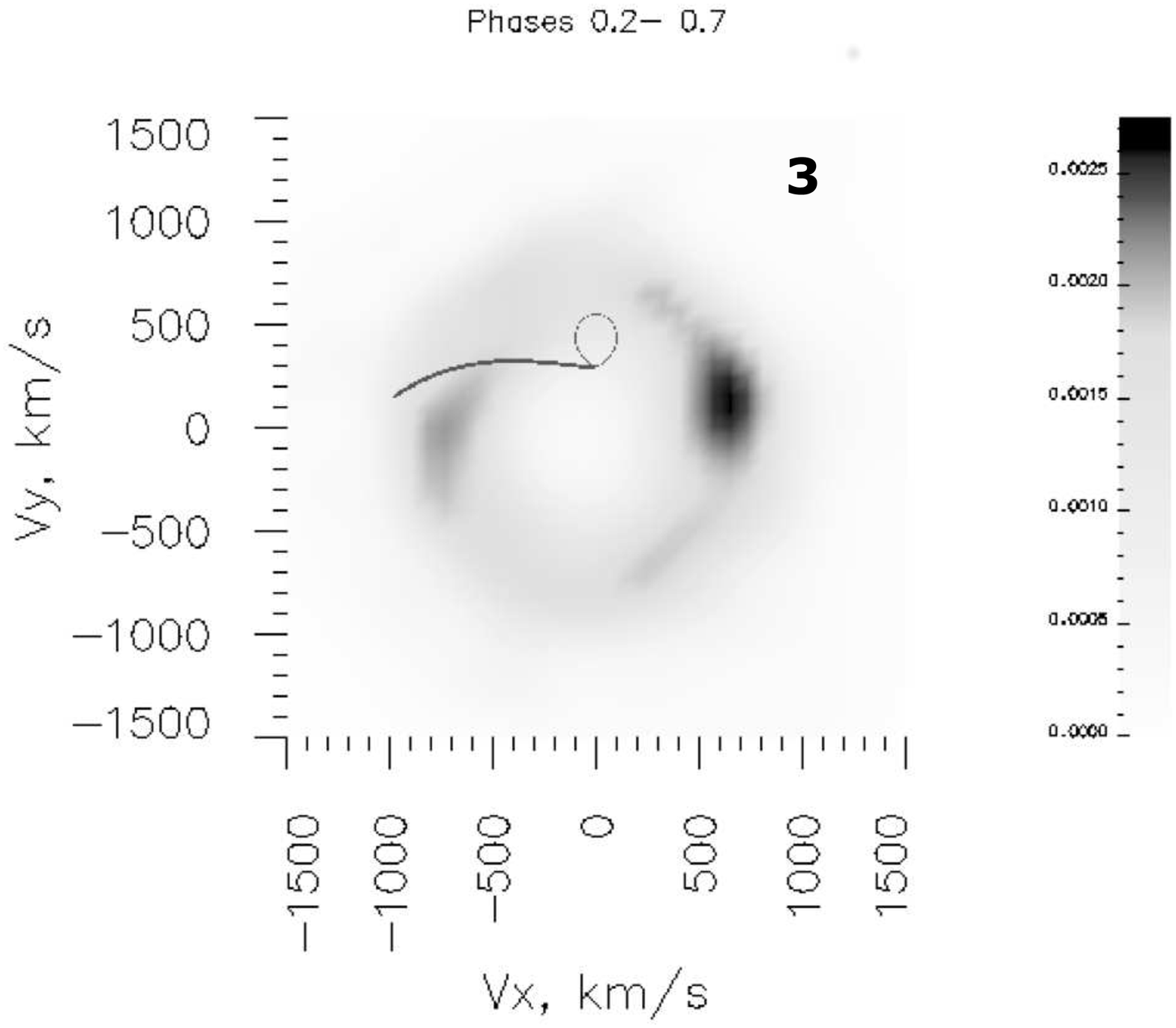}\\
\includegraphics[width=55mm]{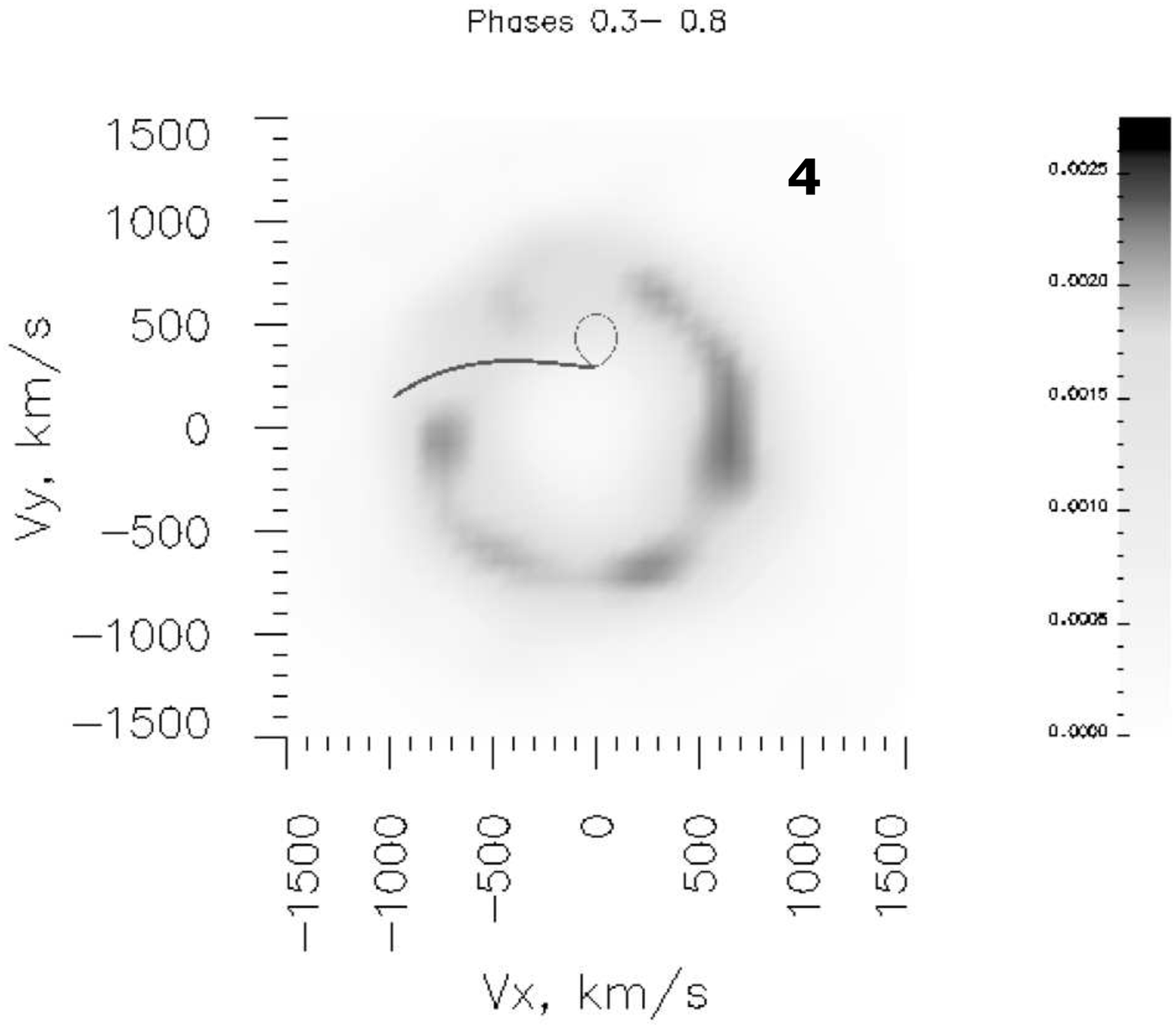}
\includegraphics[width=55mm]{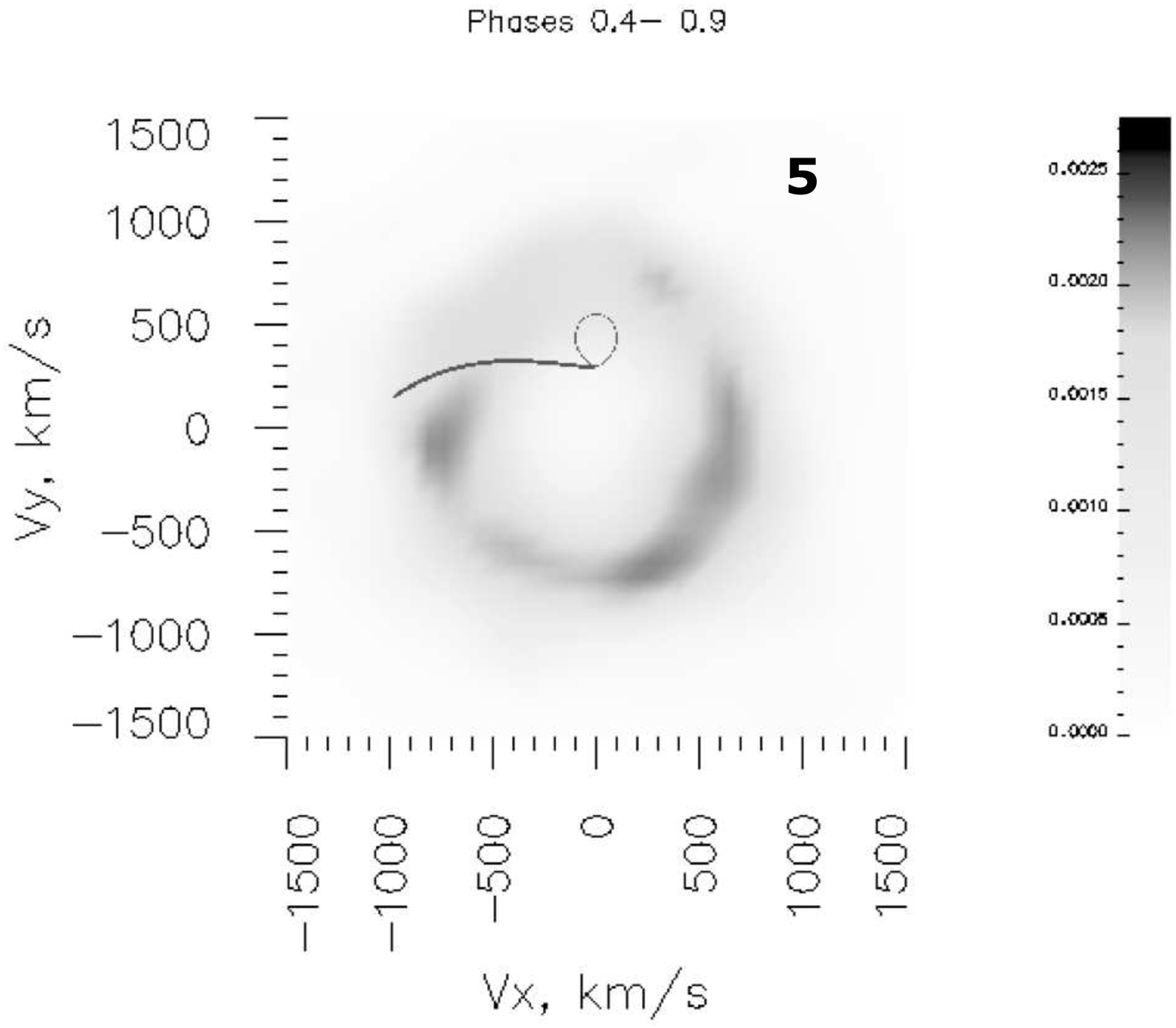}
\includegraphics[width=55mm]{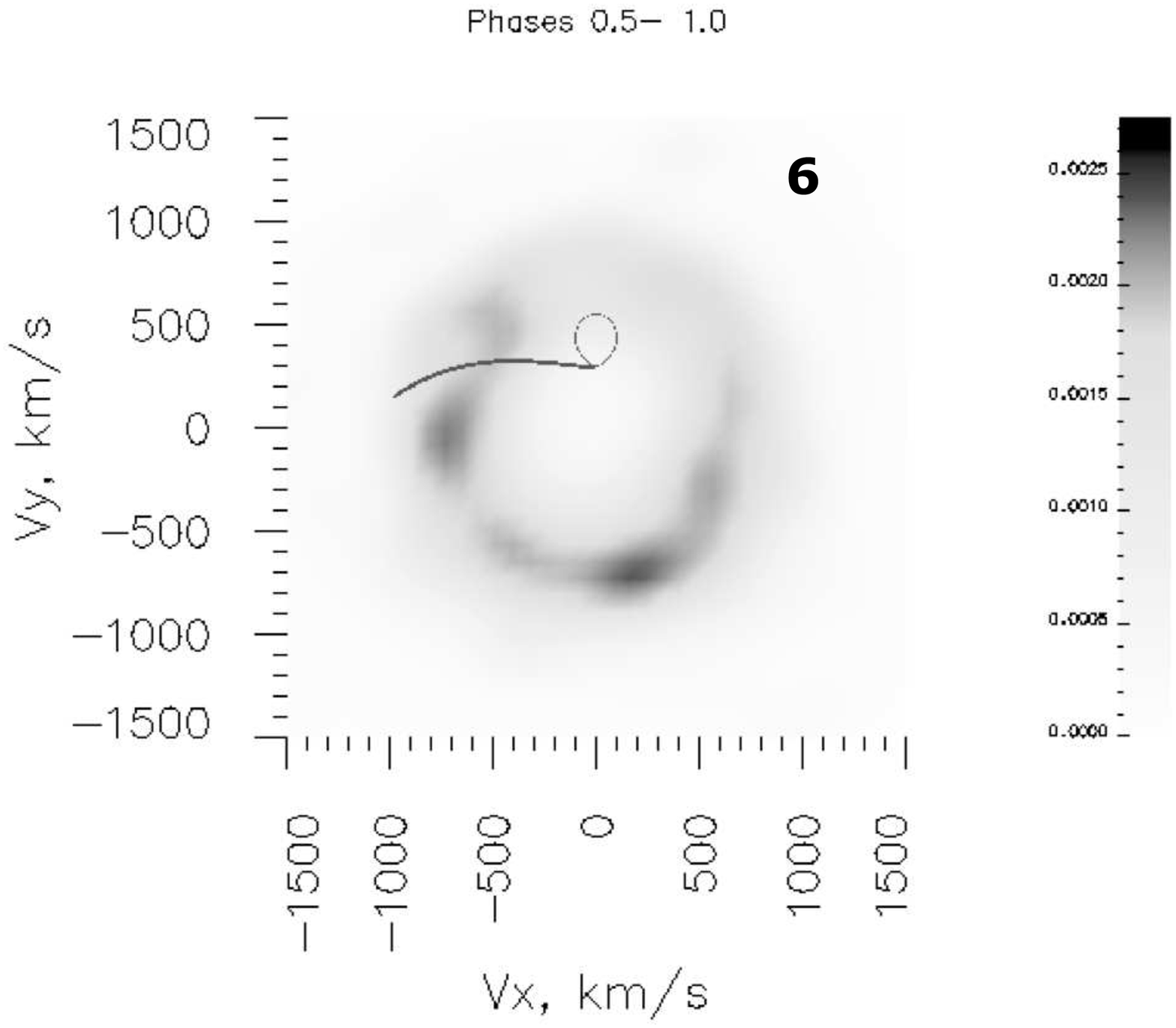}\\
\includegraphics[width=55mm]{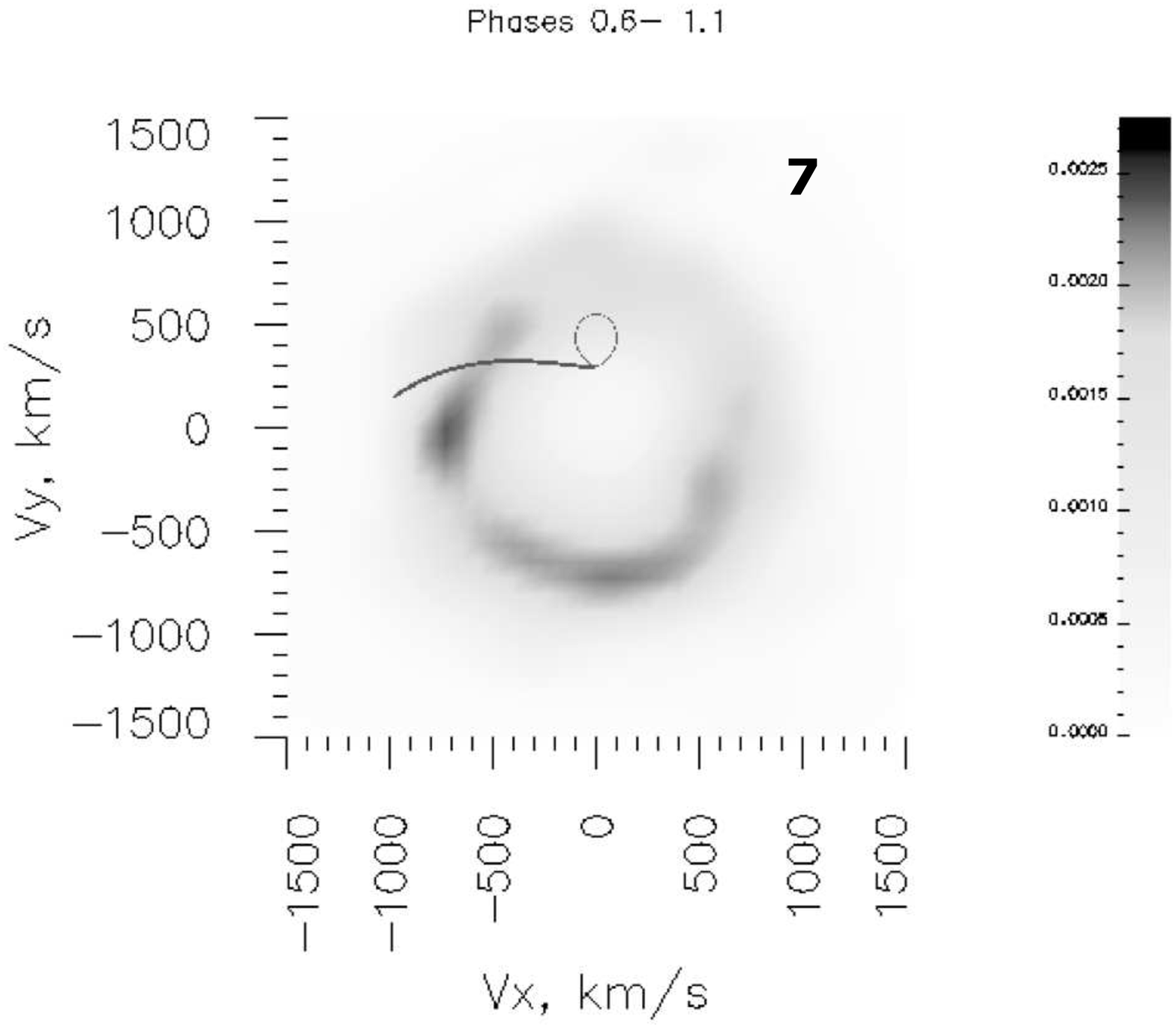}
\includegraphics[width=55mm]{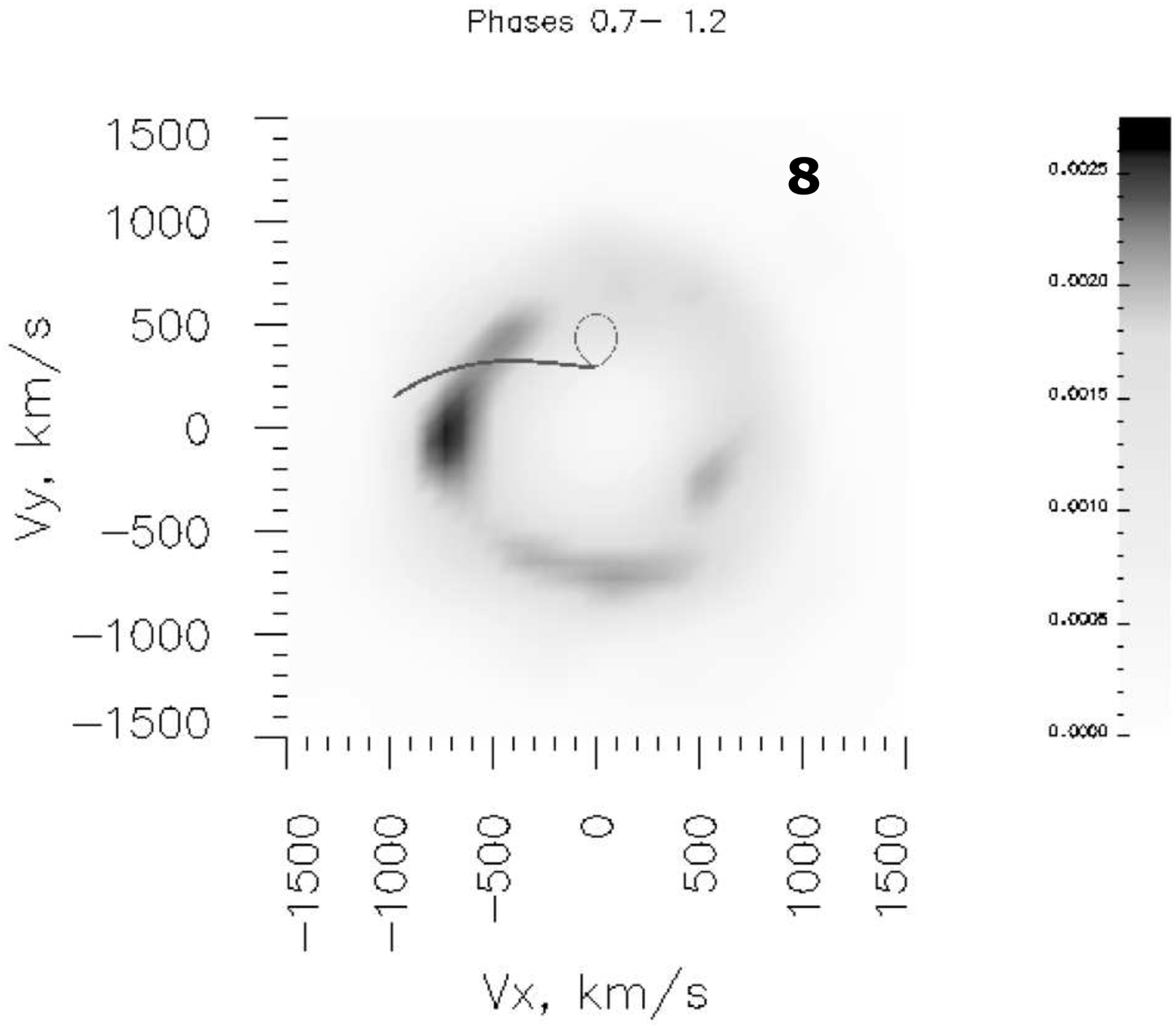}
\includegraphics[width=55mm]{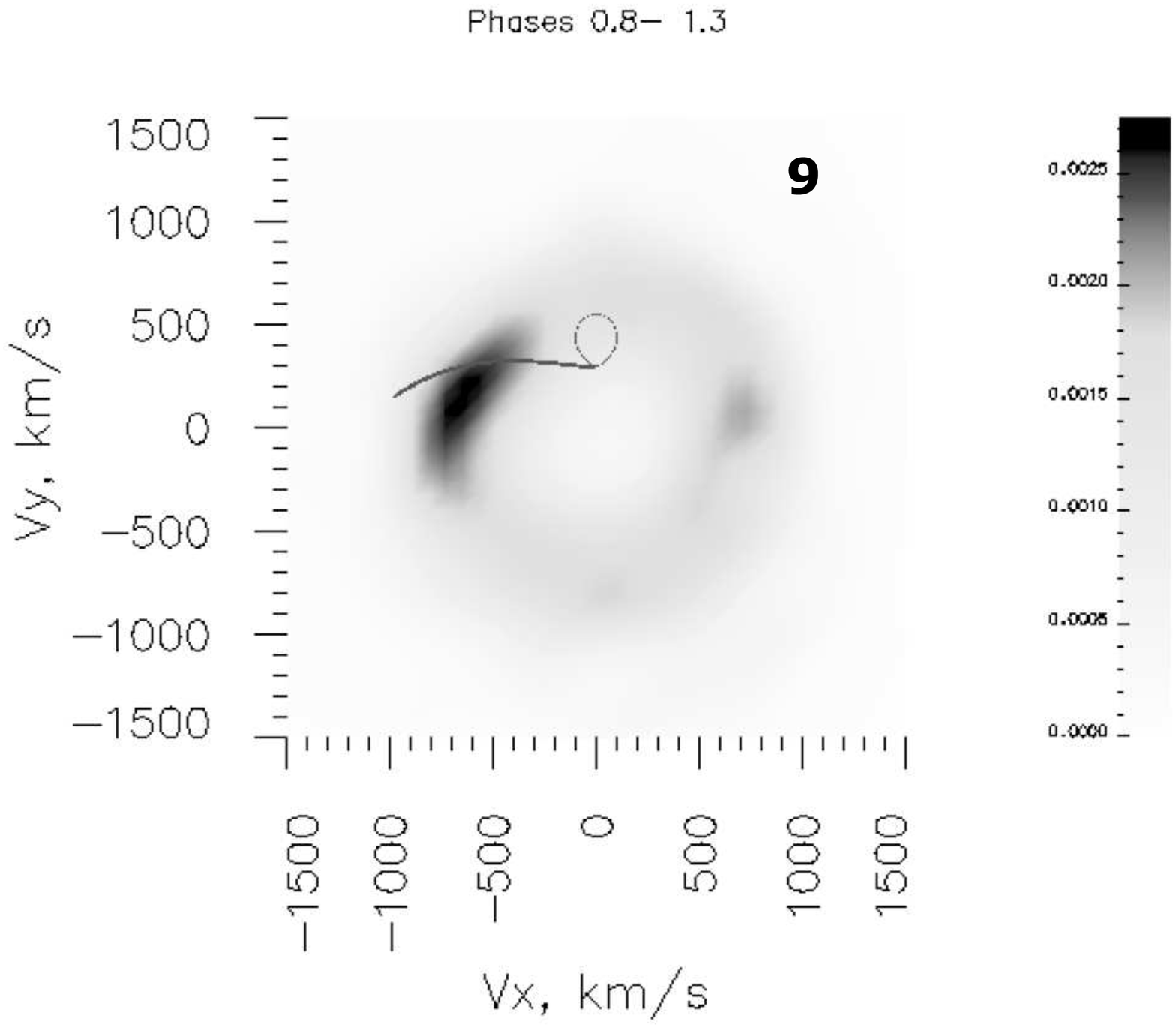}\\
\includegraphics[width=55mm]{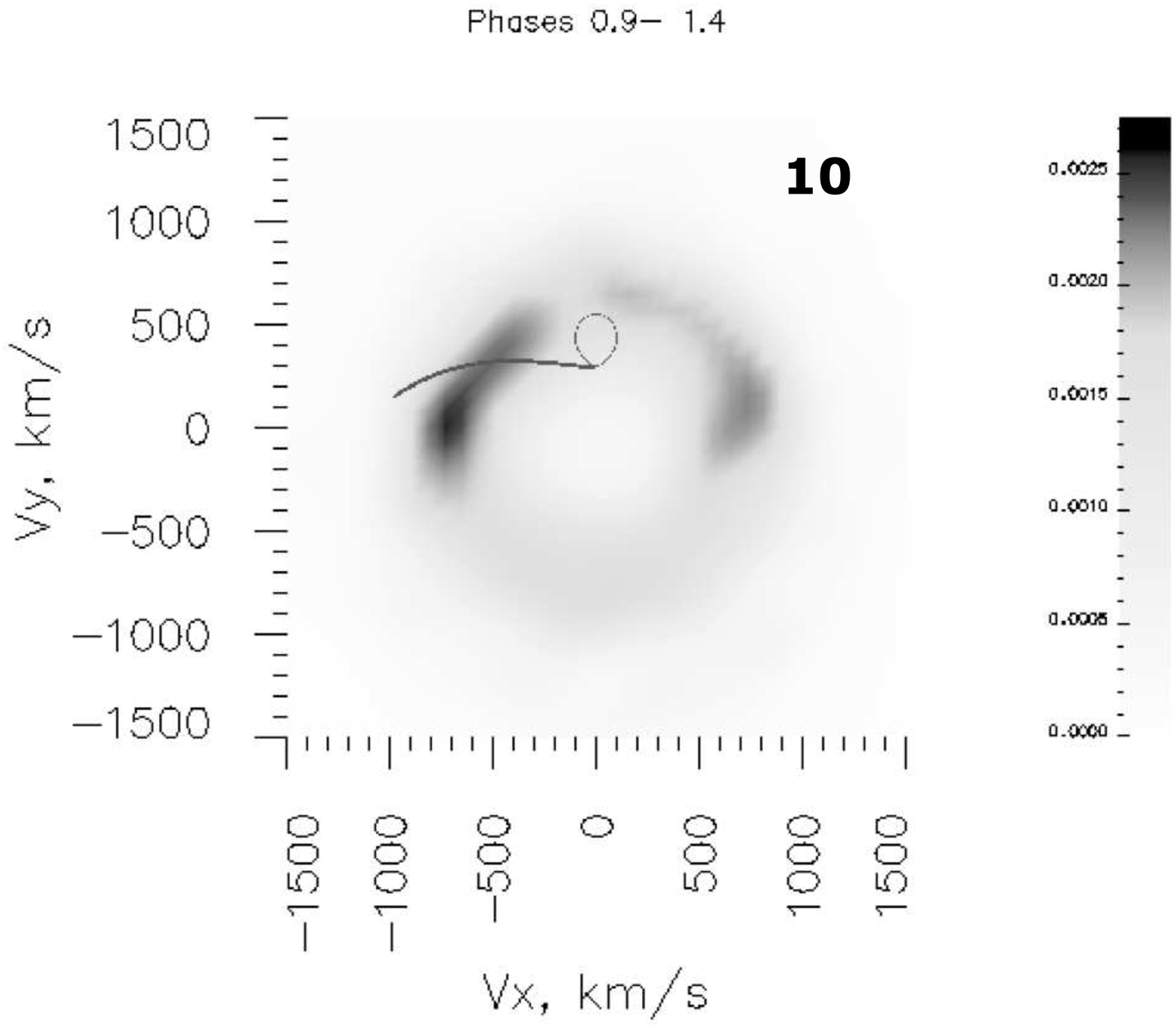}
\caption{Observational Doppler tomograms of V455~And calculated by using the spectra obtained on September 22, 2014. The intensity of each point is a fraction of unit (total intensity of the tomogram). The Roche lobe of the donor-star and the ballistic trajectory of the stream from the inner Lagrangian point $L_{1}$ in velocity coordinates are depicted by thin grey lines.}
\label{Fig6}
\end{figure*}

Note, that if one uses all the spectra obtained within a range of orbital phases of $\Delta \phi \approx 0.0 - 1.0$ to compute a tomogram, the mentioned effects of shock brightening, whose duration is shorter than the orbital period, will be smoothed out. For example, this smoothing might prevent \citet{bloem} from seeing structures other than the hot spot in the disk of V455 And. They presented the Doppler tomograms of the system in the $H_{\gamma}$, He I $\lambda4472$, and He II $\lambda4686$ and found them structureless, which led the authors to the conclusion that the disk in V455 And correspond to structureless disks of intermediate polars (IP). Meanwhile this conclusion is at least debatable, since, for example, in the results of 3D MHD numerical simulations (\citet{zhilbilA}, \citet{zhilbilB}) showed that such structures as tidal spiral shocks also exist in the disks of IPs.

The resulting tomograms are shown in Fig.~\ref{Fig6}. All the tomograms are plotted in the same intensity scale for the reader to see how they vary, depending on the range of orbital phases. One can clearly see that the tomograms are different and that the bright regions "move" over them clockwise, i.e. there is a periodicity in the brightenings.

This behavior of the tomograms, calculated using the sets of line profiles, each covering a subsequent phase interval, may be explained in two ways. First, we can consider different viewing aspects of certain features in the accretion disk. When some feature, say, a shock wave is directed toward the observer and not obscured by another features one can expect to see it brighter in observations. The other way to explain the effect, described above, is to suppose, that a particular feature can be equally visible at any angle and it becomes brighter in observation because it indeed releases a bigger amount of energy than the others at a certain moment due to physical process. For instance, this process, as we suppose in our model, may be the amplification of four major shock waves in the disk because of their interaction with the precessional density wave.

\begin{figure*}
\includegraphics[width=84mm]{fig2_a.eps}
\includegraphics[width=84mm]{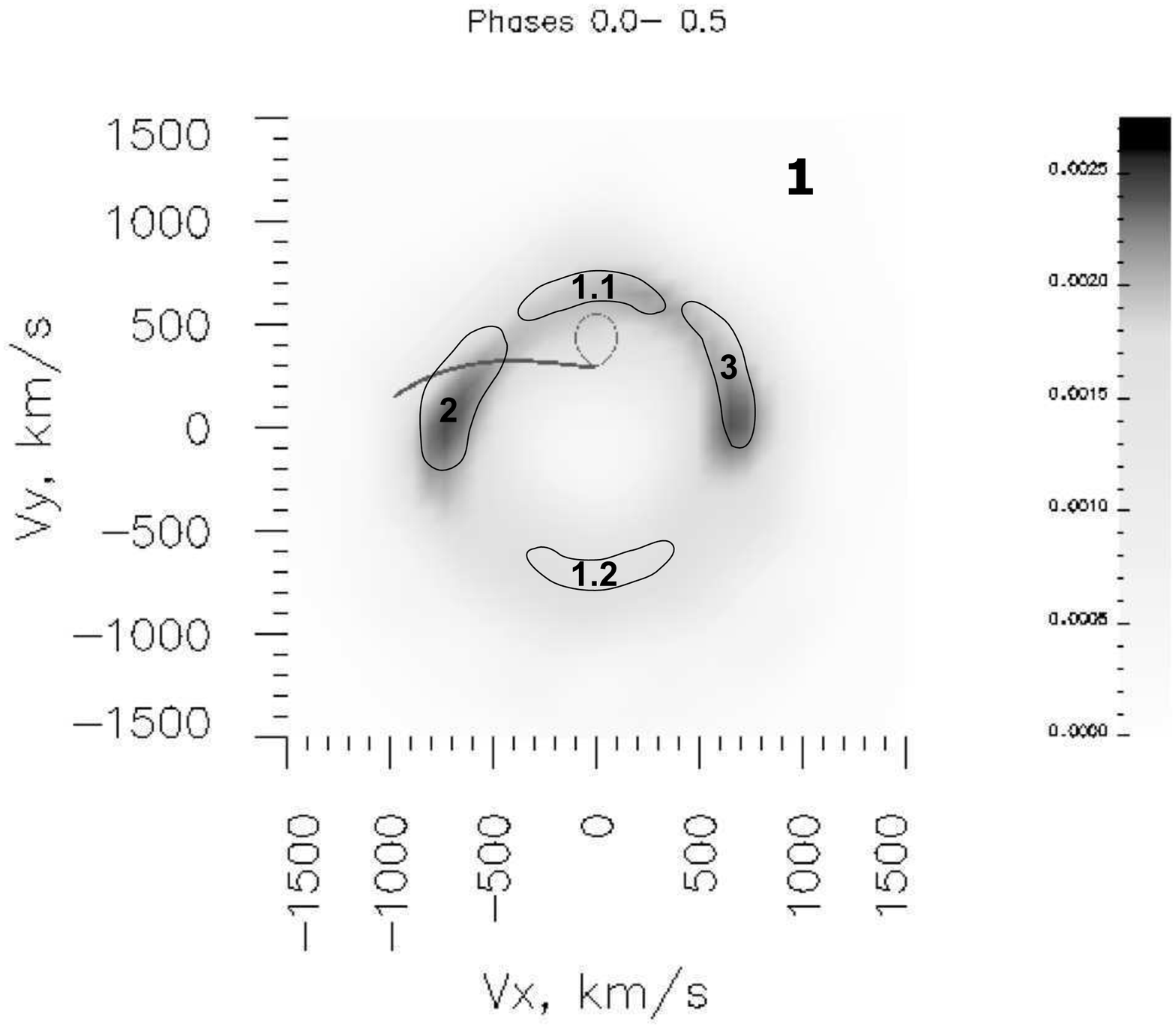}
\caption{Correspondence between the positions of the shock waves in the spatial coordinates (left-hand panel) and the tomogram (right-hand panel). The waves designated by 1.1 and 1.2 are the arms of the tidal shock wave; 2 is the hot line; 3 is bow-shock.}
\label{Fig7}
\end{figure*}

To explain where one should see each of the four mentioned shock waves in the tomogram we can consider Fig.~\ref{Fig7}. These shock waves have been identified in the tomograms of several cataclysmic variables by, e.g., \cite{shh97, KuBi01, biko08, koka08}.

\section{ANALYSIS OF THE RESULTS}
\subsection{Comparison of synthetic and observational profiles}

Since we are going to examine our model by comparing its results with observations, we have simulated possible observational manifestations, resulting from the model (synthetic line profiles and Doppler tomograms). To compute the initial synthetic line profiles we used 78 subsequent simulated gas-dynamical spatial distributions of the density, pressure, temperature, and velocity, covering a time interval from 18 to 19.36 $P_{orb}$, i.e. $\approx 1.4 P_{orb}$.  It is very important to note that, when calculating the synthetic line profiles, we did not conduct any sophisticated radiative transfer simulations. Instead, we supposed that the simulated disk (not real!) is absolutely optically thin for a spectral line. Since the emission lines, demonstrated by CVs, are believed to be recombination lines, their emissivities are proportional to the squared concentration (or, in our case, on the squared density $\rho^{2}$, see e.g. \citet{stor95}). We do not take account of the temperature, since in the simulation method we used (see \citet{ko2015}) the resulting temperature differences over the disk are small. Therefore, we suppose that the resulting emissivities depend only on the squared density, and the line profile can be expressed as follows:

$$I(V_{r}) = \int\int\int \rho^{2}(V_{x}, V_{y}, V_{z})\times g(V_{r}+V_{x}cos(\phi)sin(i)-$$
$$-V_{y}sin(\phi)sin(i)+V_{z}cos(i)) dV_{x} dV_{y} dV_{z},$$
where $V_{r}$ is the radial velocity, $V_{x}(x, y, z)$, $V_{y}(x, y, z)$, $V_{z}(x, y, z)$ are the components of gas velocity, obtained from the simulations, $\phi$ is the binary phase ($\phi=0$ corresponds to the moment $\tau\approx 18\ P_{orb}$), $i$ is the binary inclination ($i=75^{\circ}$, according to \citet{araj}), and g is the point spread function, corresponding to that of our telescope. The above expression is, of course, very simplistic, but we use it purposely. We assume the simulated disk to be absolutely transparent to avoid any effects concerned with the viewing aspects of a particular flow element. Thus, we suppose that every point in the disk is equally visible at any angle, and the resulting intensity distribution in the synthetic tomogram reflects only the actual amount of energy released at this point. In addition, we should note that our simulated disk in the quasi-stationary regime has a mean column density $\Sigma\approx0.11$ g/cm$^{2}$, which gives the vertical optical depth $\tau=\Sigma\times\kappa_{R}<1$. We estimated the Rosseland mean opacity $\kappa_{R}\approx0.0004$ by using the code of Dmitry Semenov (\citet{semen}) given that the temperature is $T\approx3000$ K, density is $\rho\approx10^{-9}$ g/cm$^{3}$, and the chemical composition corresponds to that of the Sun.

Finally, we obtained 78 line profiles, each corresponding to a momentary distribution of physical parameters in the disk. To phase-bin these synthetic profiles we assigned $\phi = 0.0$ to the simulated momentary distribution where the brightest shock wave is the hot line (see top-left panel of Fig.~\ref{Fig2}). This approximately corresponds to the phase binning of the observed spectra. One can see that the brightest region, corresponding to the hot line, appears in the observational tomograms (Fig.~\ref{Fig6}.) calculated by using profiles within phase intervals $\Delta \phi \approx 0.0 - 0.5; 0.7 - 1.2; 0.8 - 1.3; ... 0.9 - 1.4$. The synthetic trailed spectra are shown in the right-hand panel of Fig.~\ref{Fig5}. The central faint S-wave observed in the simulated spectra is caused by the dense part of the stream issuing from the inner Lagrangian point.

\begin{figure*}
\includegraphics[width=140mm]{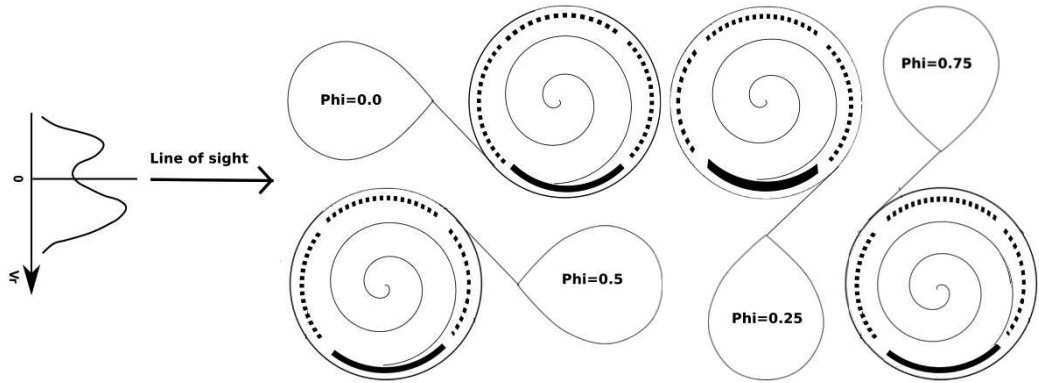}
\caption{Schematic view of the line profile formation. The teardrop shape represents the mass-losing secondary star in the system. The solid-line arcs depict currently amplified shock waves.}
\label{Fig8}
\end{figure*}

By considering the observed and simulated trailed spectra we can see that they have similar features, though the observed spectra are quite noisy. First, one can see that the simulated and observed spectra have approximately same width. Then in both the panels we observe the asymmetry of the red and blue humps of the profile. The red hump of the simulated spectra is brighter than the blue hump. The red hump of the observed spectra is not much brighter than the blue hump, but it's somewhat broader. This behavior of both the simulated and observed profiles can be explained as follows. An ideal disk with uniformly distributed brightness should give double-humped line profiles with equally high and broad humps (see, e.g. \citet{Marsh88}). In our case, in the simulated data at the orbital phase $\phi=0.0$ we observe the bright region of the hot line (see top-left panel of Fig.~\ref{Fig2} and Fig.~\ref{Fig8} ). This bright region occurs due to the amplification of the hot line by the precessional density wave and contributes to the red hump of the corresponding line profile, making it higher and somewhat broader (see Fig.~\ref{Fig5}, right-hand panel). As we mentioned above, the precessional angular velocity differs from the orbital angular velocity by only 1-2\%. Thus, the precessional density wave almost rests in the observer's coordinate frame within one orbital period. This means that the consequently amplified shock waves should be seen at almost the same place in the observer's coordinate frame and contribute to the same hump of the double-humped profile within one orbital period (see Fig.~\ref{Fig8}). We clearly see this effect in the simulated trailed spectra where the red hump is always brighter. In the observed spectra we can see that the red hump is always broader than the blue one. Probably, this may be explained by the same effect, but the difference between the intensity of the corresponding amplified shock and the rest of the disk may be smaller and we cannot see it well in the observed profiles.

It is interesting to note that in Fig. 10 of the paper of \citet{araj} the $H_{\beta}$ trailed spectra demonstrate almost the same behavior as our simulated profiles, but in that paper the blue hump of the line profile is always brighter. This may be explained by the different position of the density wave in the observer's coordinate frame, during their observations. At the same time high-quality spectroscopic data on V455 And reported by \citet{bloem} exactly correspond to our synthetic trailed spectra. In the $H_{\gamma}$, He I $\lambda4472$, and He II $\lambda4686$ trailed spectra presented by \citet{bloem} one can clearly see that within the whole observational interval the red hump of all the three lines was stronger than the blue hump, as seen in our model data, too. \citet{bloem} even make a proposition that this behavior of their data may be in some way related to the precession of the disk. These observational facts also strongly support our model.

Thus, by noting the mentioned similarity of the simulated and observed trailed spectra, we may suppose that the optical depth of the real disk in V455 And in the observer's direction is also small, since the simulated spectra were obtained under the assumption that the simulated disk is optically thin. It is interesting to note that an optically thin disk in WZ Sge was earlier proposed by \citet{mason2000}. This fact is important for us, since it supports our idea that the behavior of the observational Doppler tomograms is determined by physical processes but not by the viewing aspects of certain structures.

\subsection{Comparison of synthetic and observational tomograms}

\begin{figure*}
\includegraphics[width=55mm]{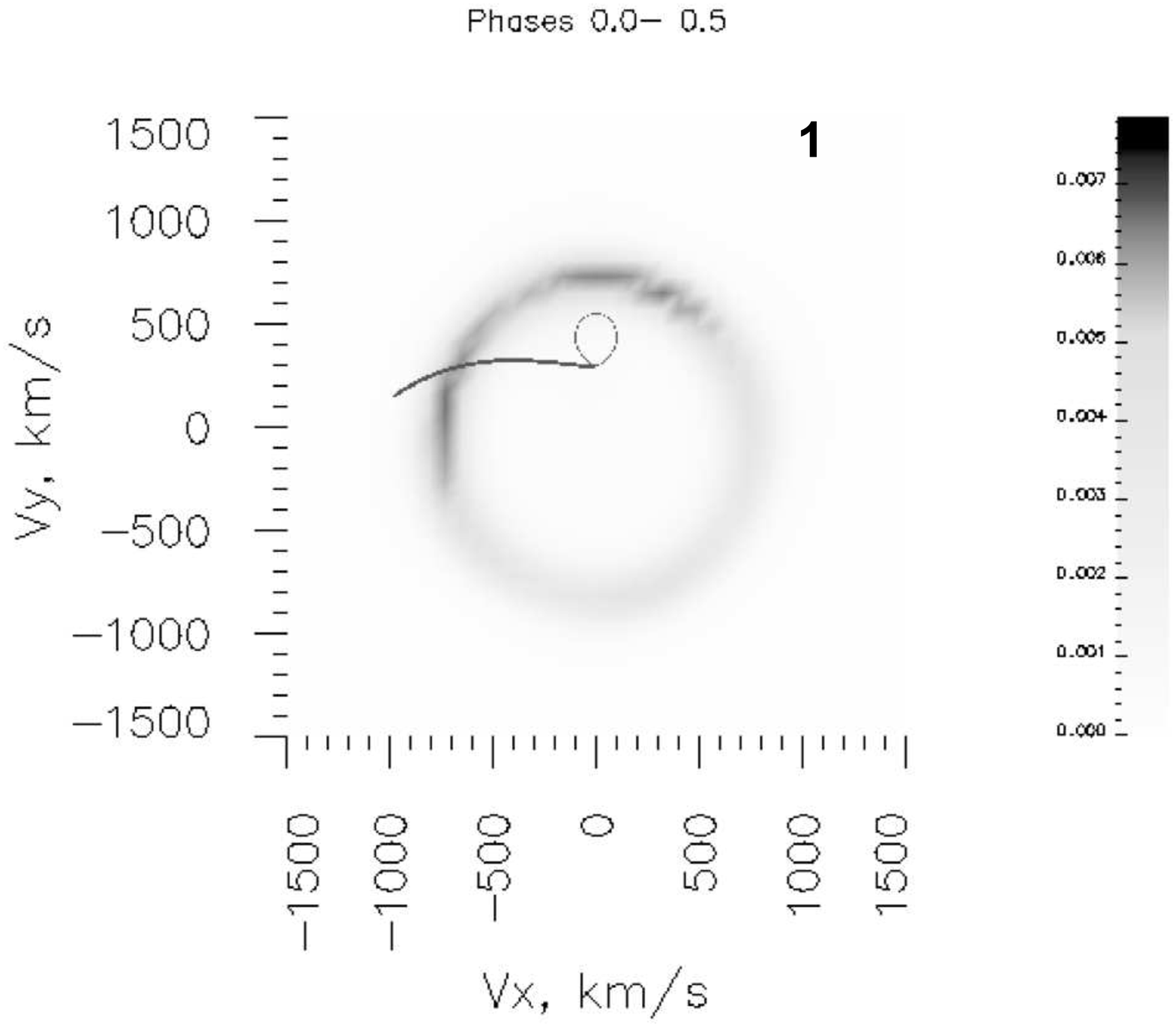}
\includegraphics[width=55mm]{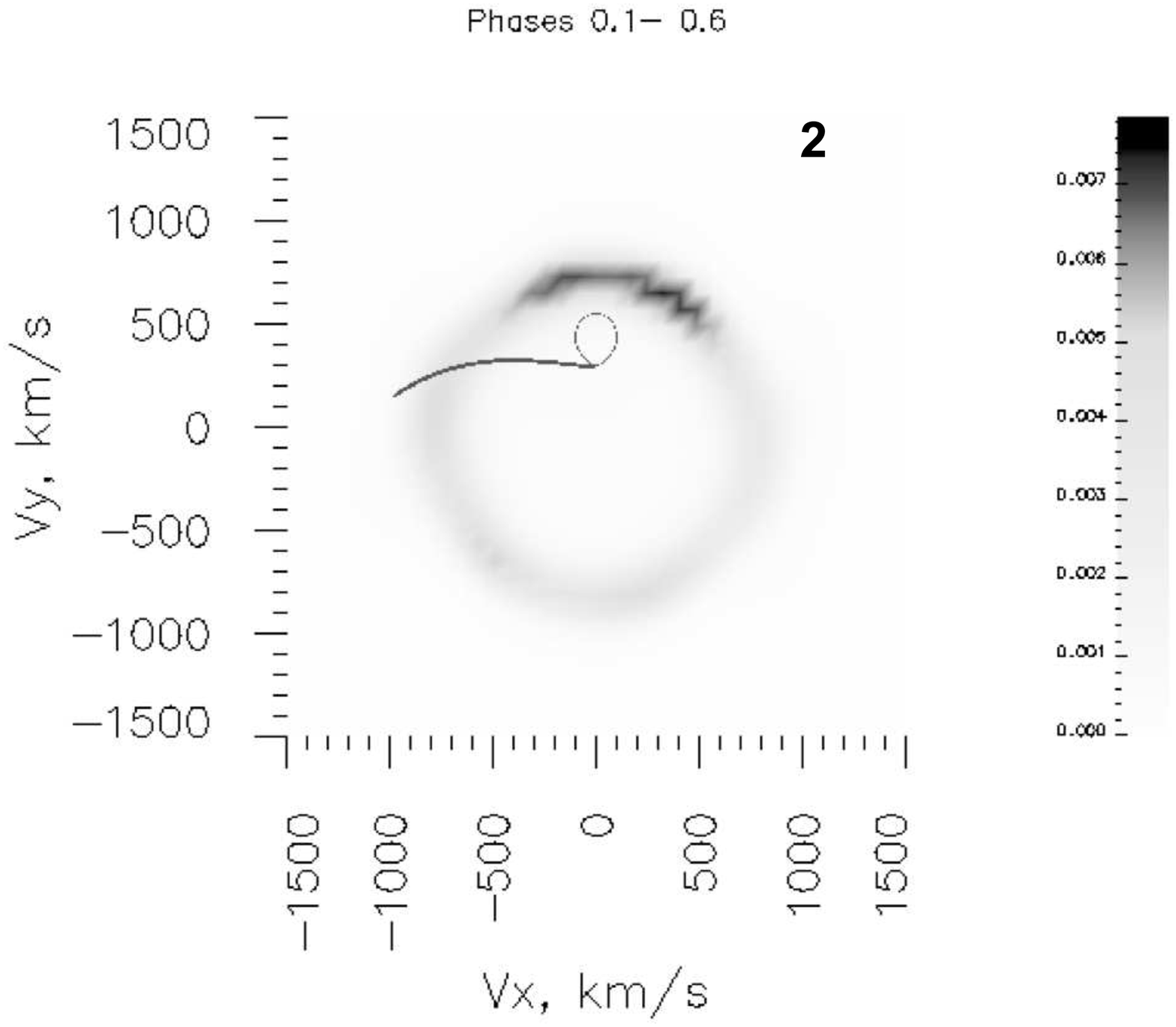}
\includegraphics[width=55mm]{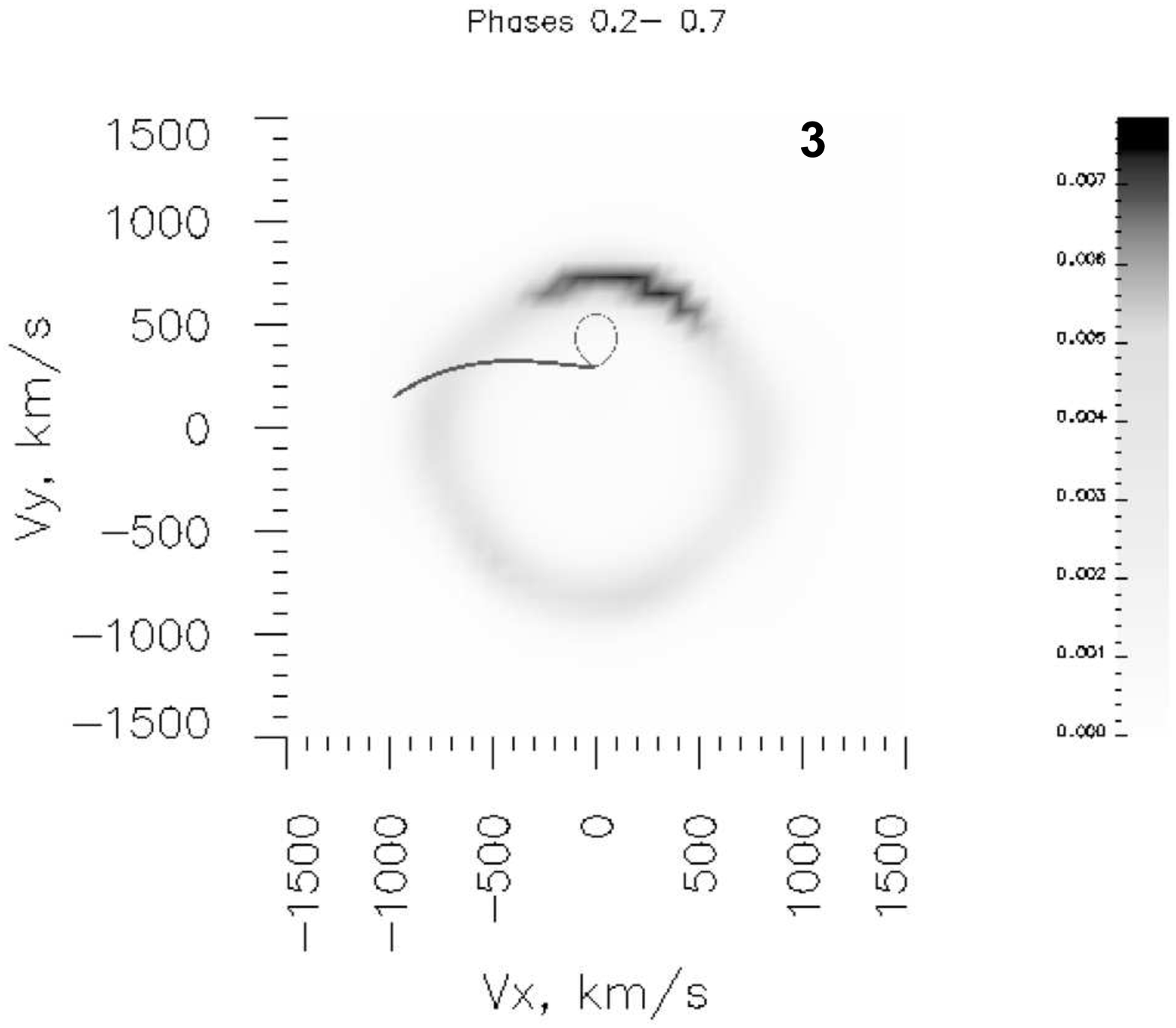}\\
\includegraphics[width=55mm]{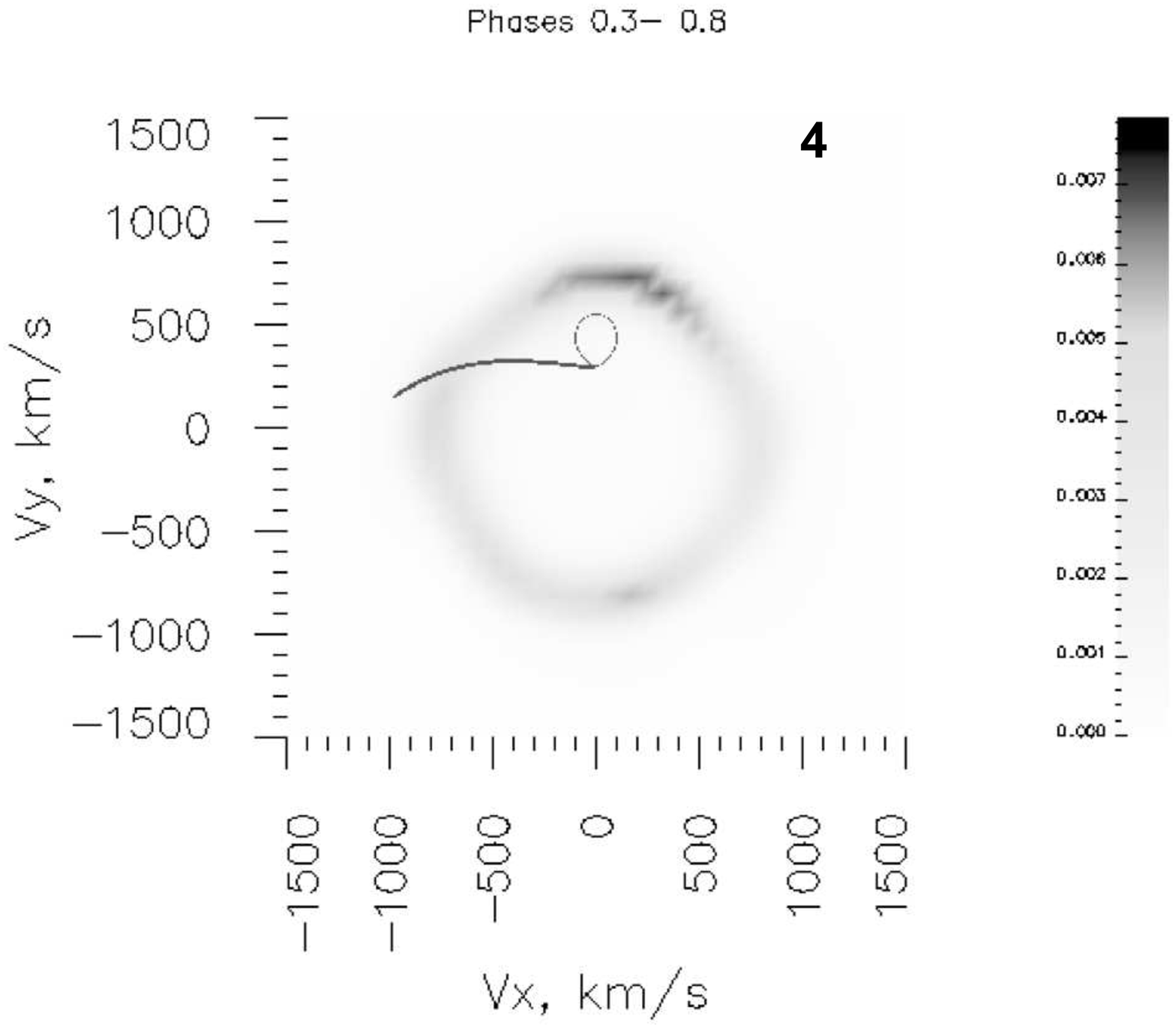}
\includegraphics[width=55mm]{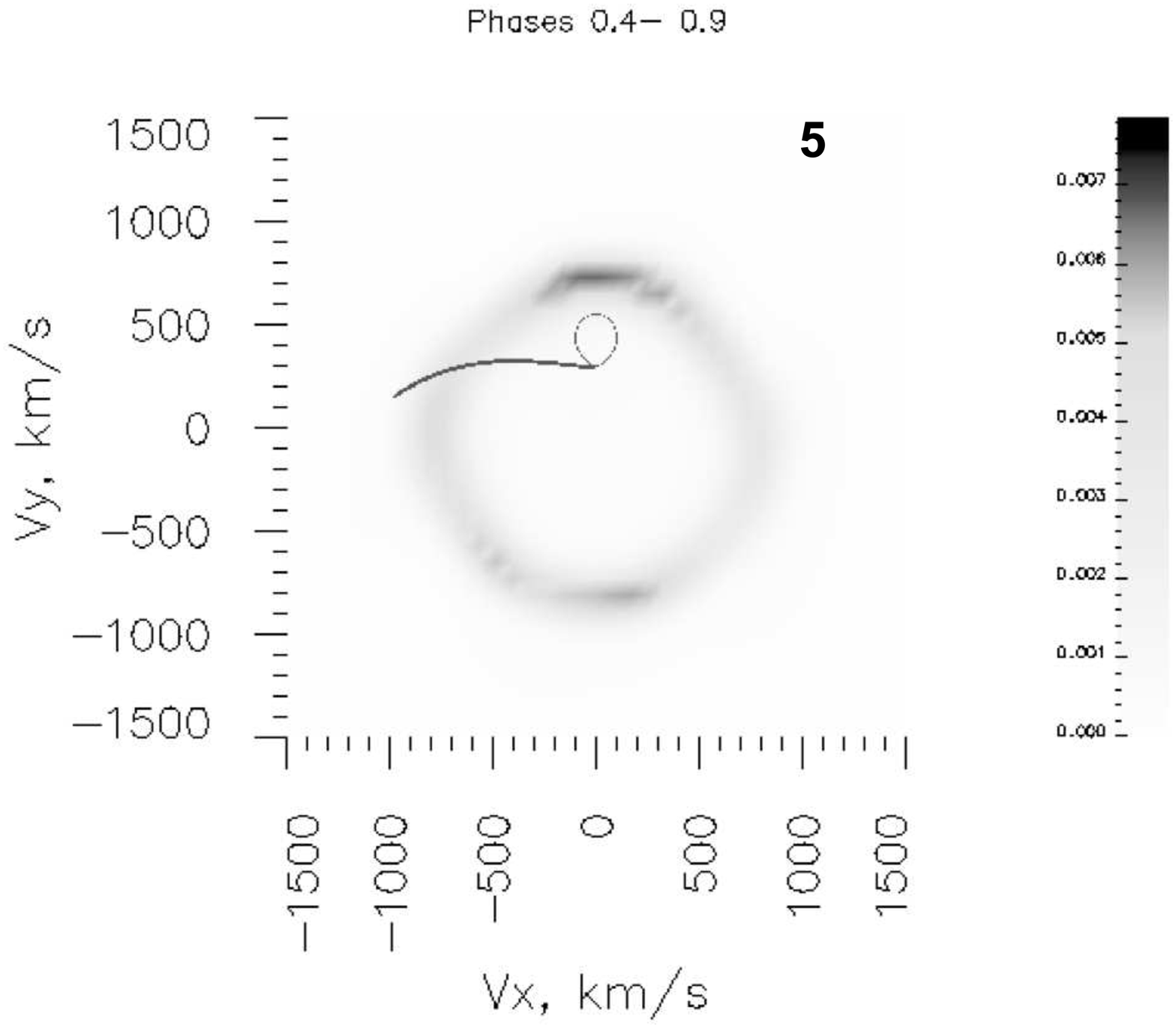}
\includegraphics[width=55mm]{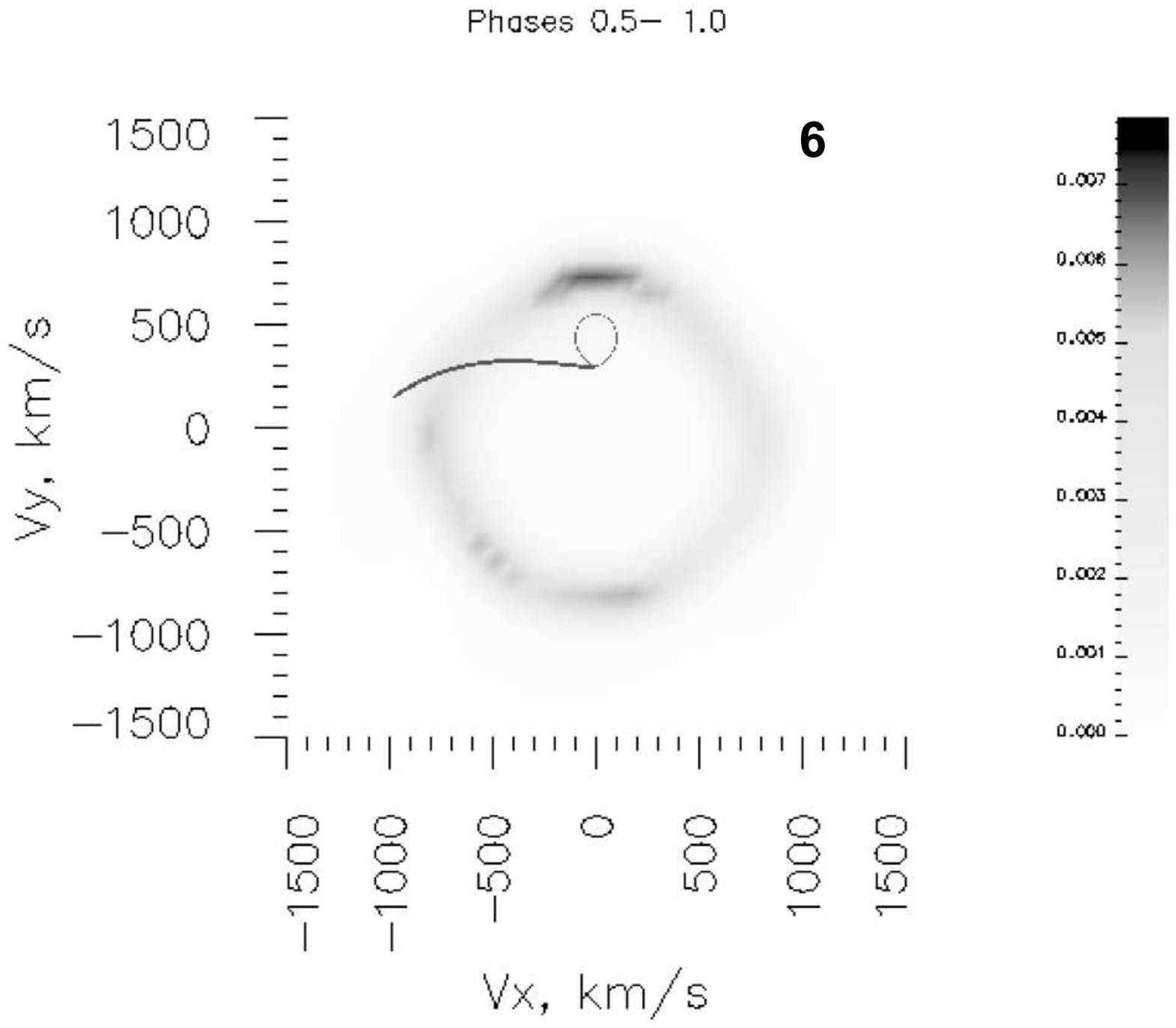}\\
\includegraphics[width=55mm]{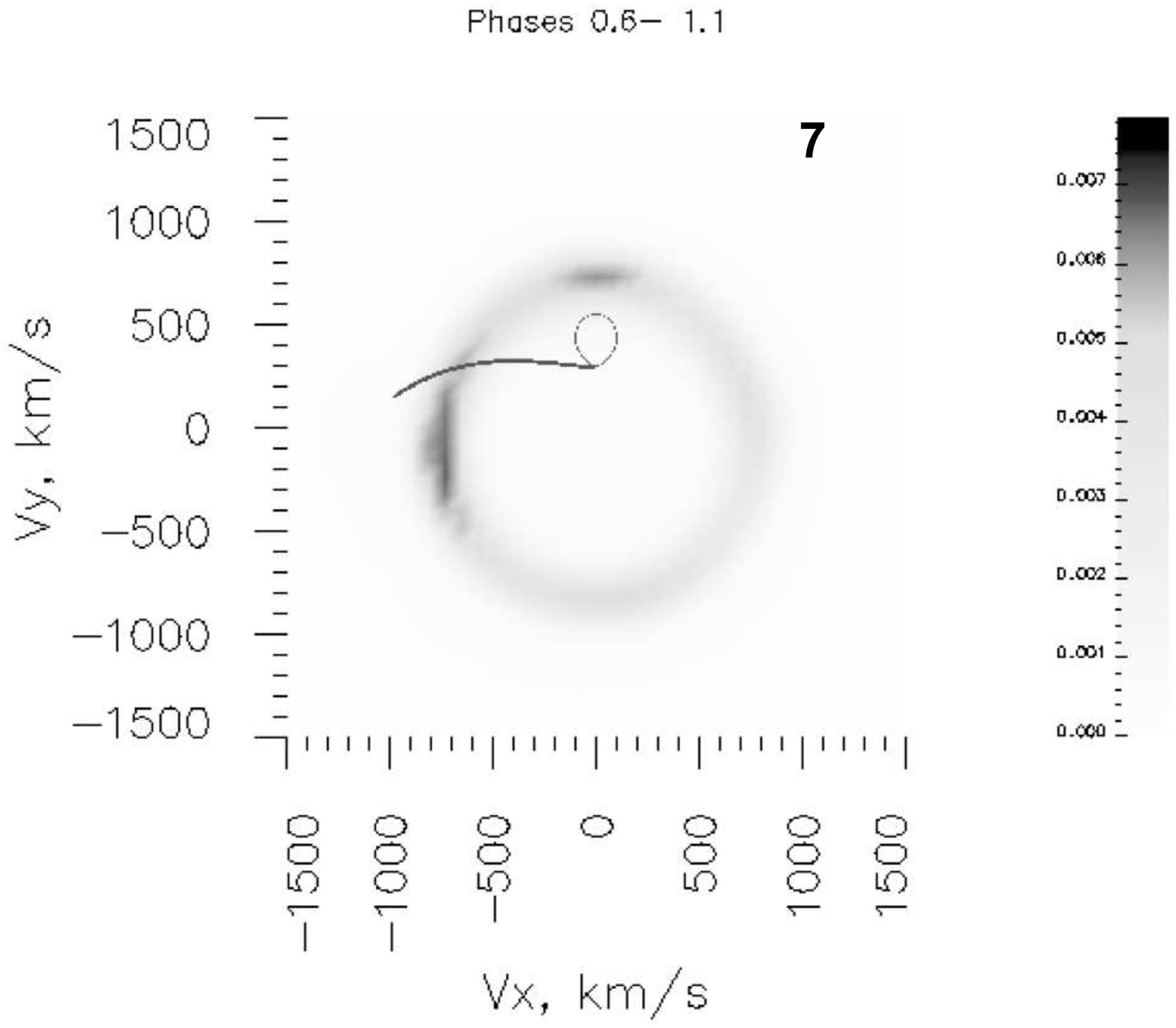}
\includegraphics[width=55mm]{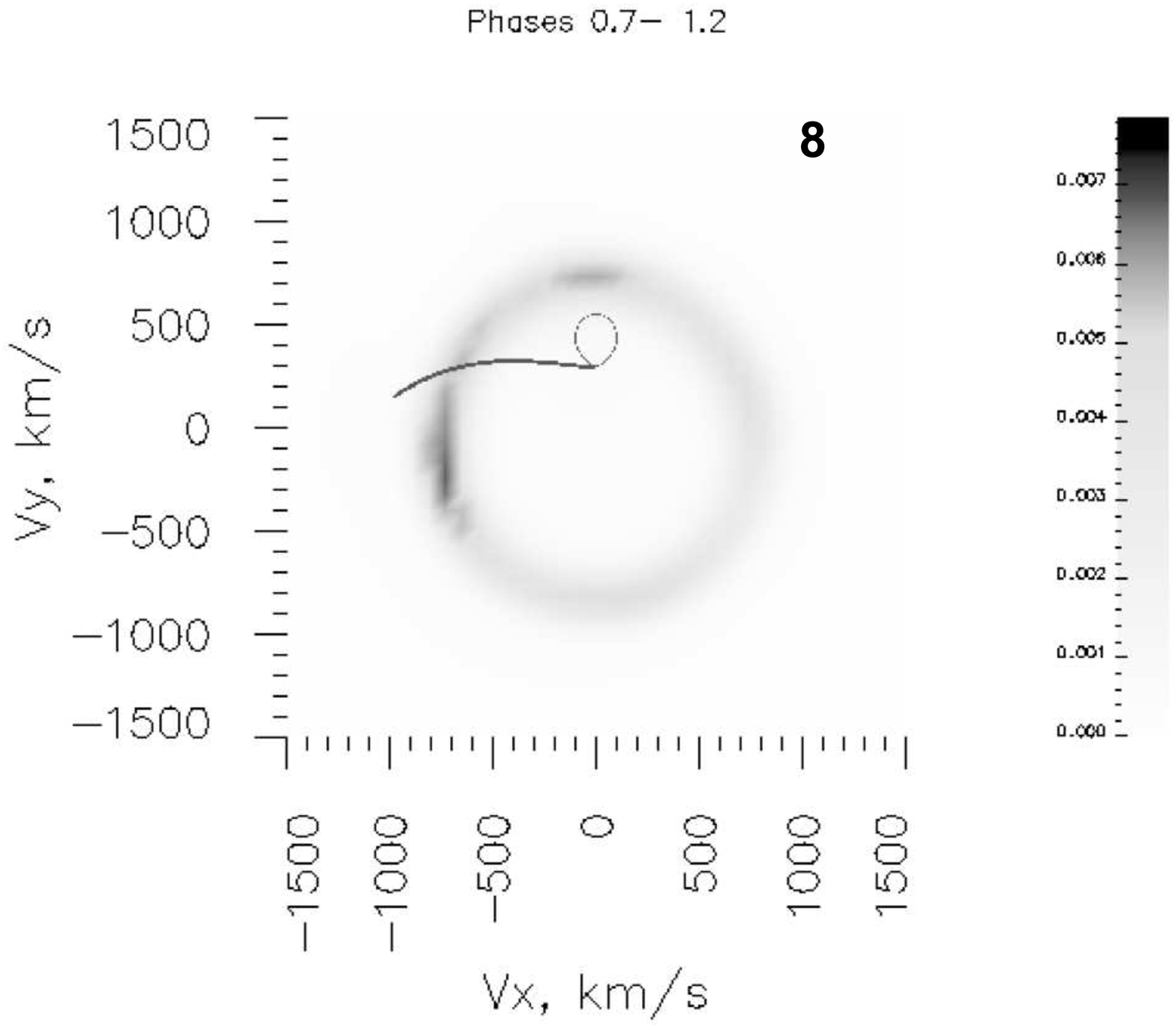}
\includegraphics[width=55mm]{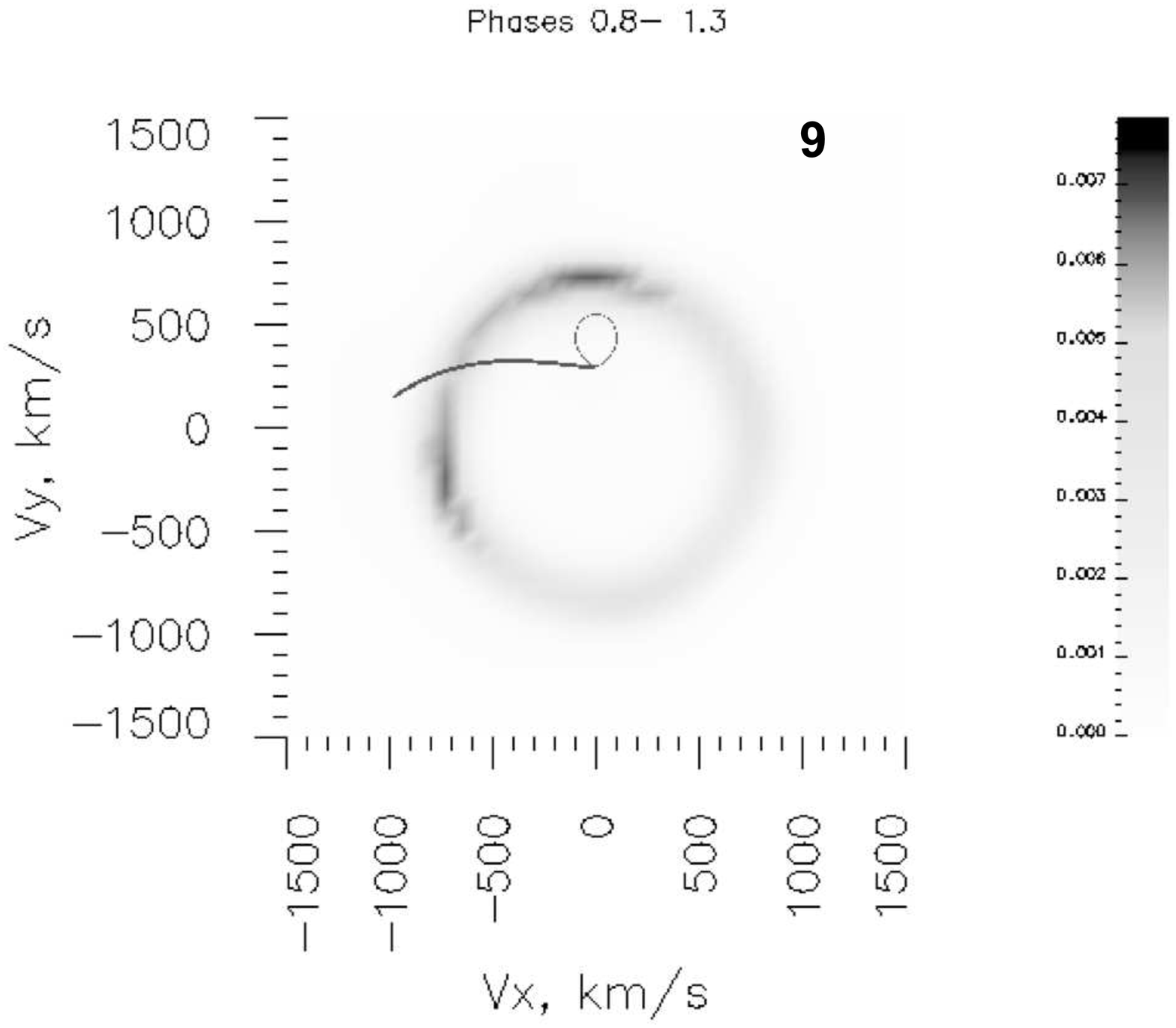}\\
\includegraphics[width=55mm]{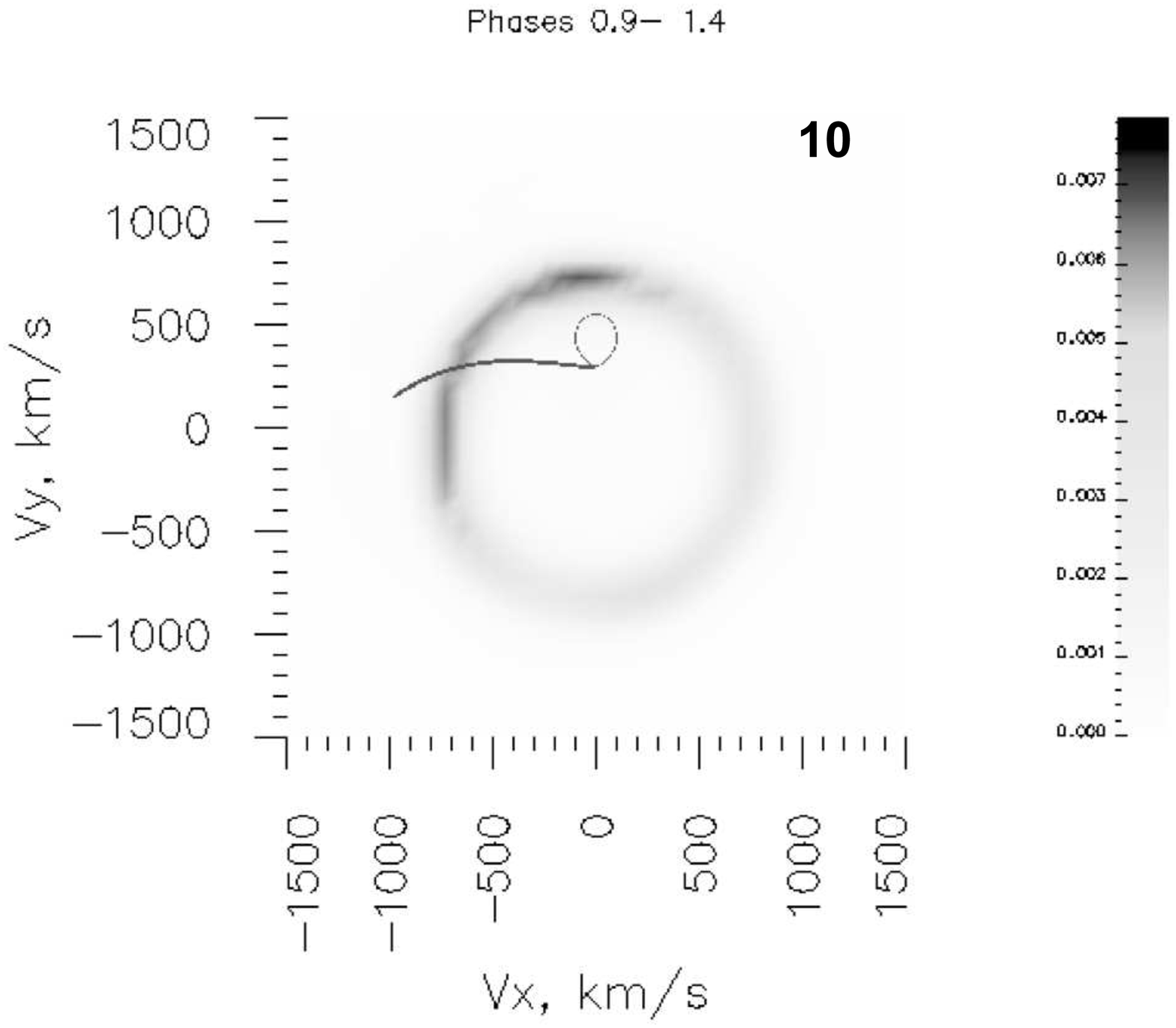}
\caption{Synthetic Doppler tomograms of V455~And calculated by using the synthetic emission line profiles in the same orbital phase intervals as in Fig.~\ref{Fig6}. The intensity of each point is a fraction of unit (total intensity of the tomogram). The Roche lobe of the donor-star and the ballistic trajectory of the stream from the inner Lagrangian point $L_{1}$ in the velocity coordinates are depicted by thin grey lines.}
\label{Fig9}
\end{figure*}

To calculate the synthetic Doppler tomograms we also subdivided the entire sample of simulated profiles into 10 subsamples: $\phi \approx0.0 - 0.5$, $0.1 - 0.6$, $0.2 - 0.7$...$0.9 - 1.4$. The resulting synthetic Doppler tomograms of the V455~And system are shown in Fig.~\ref{Fig9}. They are again plotted in the same intensity scale for the reader to see how they vary with time. The bright region in the synthetic tomograms also shifts clockwise from one tomogram to another. Since we know the initial distribution of velocity in our simulations, we can unambiguously attribute each bright region in the synthetic tomograms to features in the simulated accretion disk. In our case these regions exactly correspond to the shock waves in the disk (see Fig.~\ref{Fig2} and Fig.~\ref{Fig7}). Since we suppose the simulated disk to be absolutely transparent, we can state that the behavior of the synthetic tomograms clearly traces the process of the shock wave amplification.

\begin{figure*}
\includegraphics[width=84mm]{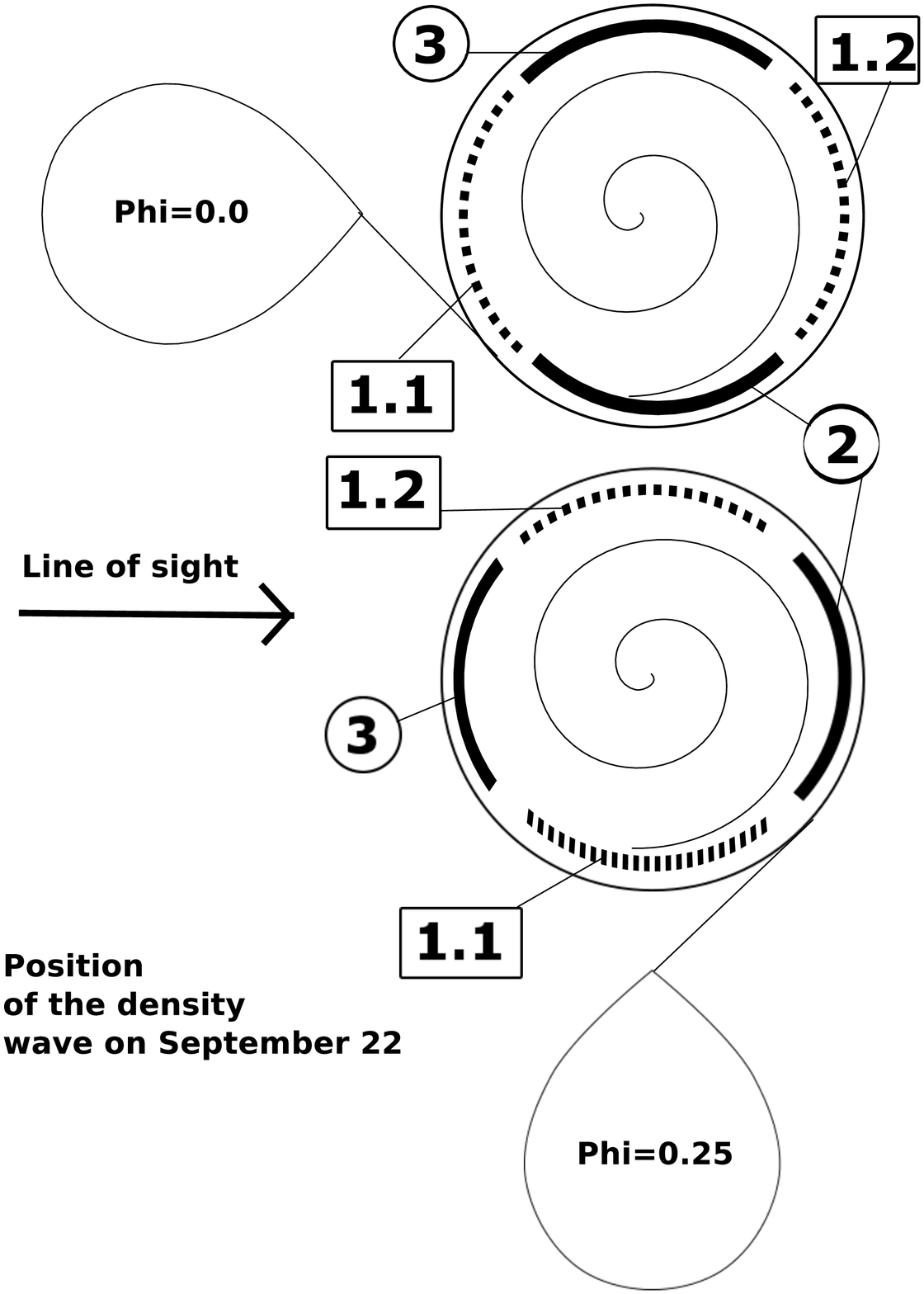}
\includegraphics[width=80mm]{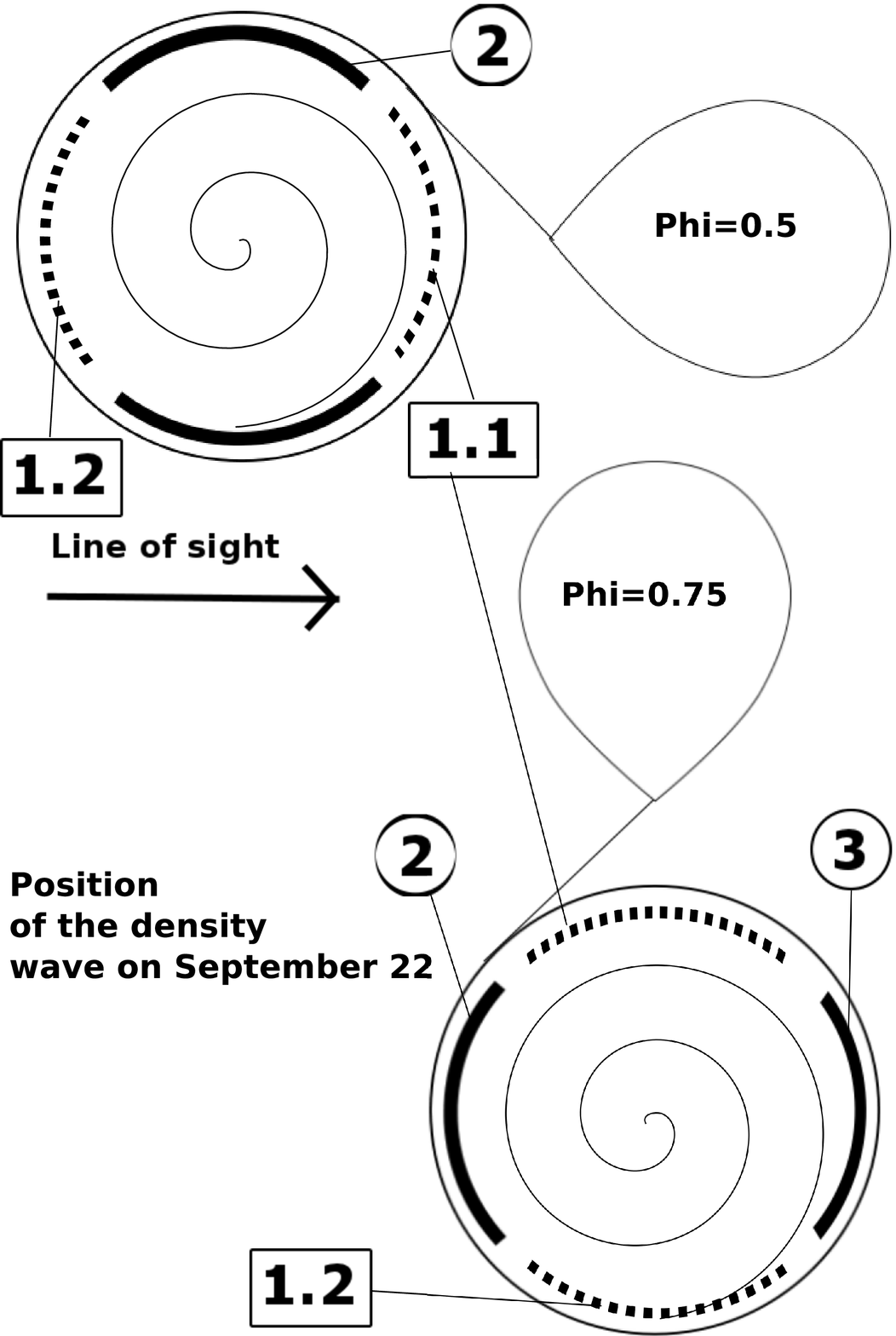}
\caption{Positions of major features in the binary as a function of orbital phase (Phi) for four different values of Phi. The teardrop shape represents the mass-losing secondary star in the system. The orbital plane of V455~And actually does not lay exactly along the observer's line of sight. The actual binary inclination is $i\approx75^{\circ}$ (\citet{araj}). The shock waves are designated by numbers in the same way as in Fig.~\ref{Fig1}.}
\label{Fig10}
\end{figure*}

It is obvious that the synthetic tomograms look and behave similarly to the observational tomograms. Even the bright regions in both of the results are located at similar velocity coordinates (compare Fig.~\ref{Fig6} and Fig.~\ref{Fig9}). This allows us to suppose that the behavior of the observational tomograms is also determined by physical processes in the accretion disk instead of viewing aspects of a particular feature. There is another fact that supports our idea. We see that the brightest regions occur in the observational tomograms, calculated using the spectra within the phase intervals $\Delta \phi \approx 0.2 - 0.7$ (tomogram 3, Fig.~\ref{Fig6}, right-hand quadrants) and $\Delta \phi \approx 0.8 - 1.3$ (tomogram 9, Fig.~\ref{Fig6}, left-hand quadrants). As one can see in Fig.~\ref{Fig7} these two regions in the tomograms are attributed to the bow-shock (in the right-hand quadrants of a tomogram) and the hot line (in the left-hand or top-left quadrants of a tomogram). However, as one can see on the scheme in Fig.~\ref{Fig10} these shock waves, designated by numbers $\textcircled{2}$ (hot line) and $\textcircled{3}$ (bow-shock), are exposed directly to the observer when the system is observed at orbital phases near 0.2 - 0.25 and 0.75 - 0.8. Hence, if the brightness of these waves in the tomogram is determined only by the viewing aspects we should expect to see the maximal intensities within the phase intervals  $\Delta \phi \approx 0.0 - 0.5$ (bow-shock) and  $\Delta \phi \approx 0.5 - 1.0$ (hot line), where these waves are directly exposed to us. However, we see them becoming brighter at least $0.2 P_{orb}$ later than when they pass the line of sight.

Summarizing the results, we can propose the following explanation for the behavior of the observational tomograms of September 22. We, as usual, suppose that the binary system rotates counterclockwise and the density wave, as mentioned above, almost rests in the observer's coordinate frame. When the system is observed near the phase $\phi=0.25$ (the midpoint of the interval $\Delta \phi \approx 0.0 - 0.5$), the outer part of the density wave causes a density increase near the arm of the tidal shock located closer to the donor-star (left-hand panel of Fig.~\ref{Fig10}, bottom scheme), which is observed as the bright region in the upper quadrants of tomogram 1 (Fig.~\ref{Fig6}) above the Roche lobe of the donor-star. In tomogram 2 of Fig.~\ref{Fig6} we see two bright regions in the left and right quadrants associated with the hot line and bow-shock, respectively. It is interesting to note that this tomogram is calculated by using the spectra in the phase interval $\Delta \phi \approx 0.1 - 0.6$, within which the hot line is mostly located in the opposite side of the accretion disk. Nonetheless we still see it almost as bright as the bow-shock directly exposed to us. This again allows us to suppose that the position of the shock waves and their viewing aspects play a smaller role in the intensity distribution of the tomograms.

Within the phase interval $\Delta \phi \approx 0.2 - 0.7$ (the midpoint is $\phi \approx 0.45$) we see the amplification of the bow-shock (right-hand panel of Fig.~\ref{Fig10}, top scheme) contributing to the right-hand quadrants of tomogram 3 (Fig.~\ref{Fig6}). In tomograms 4 and 5 (Fig.~\ref{Fig6}) we see how the bow-shock passes through the outer part of the density wave. Near the phase $\phi=0.75$ (the midpoint of the interval $\Delta \phi \approx 0.5 - 1.0$) the density wave interacts with the second arm of the tidal shock (right-hand panel of Fig.~\ref{Fig10}, bottom scheme), which gives the bright spot in tomogram 6 of Fig.~\ref{Fig6} in bottom quadrants. When we see the system at the phase $\phi\approx1.0$ (the midpoint of the interval $\Delta \phi \approx 0.8 - 1.3$), the outer part of the density wave is located near the hot line (left-hand panel of Fig.~\ref{Fig10}, top scheme). This  results in the local increase of density near the shock wave and, hence, in its amplification and brightening, which is seen in the left-hand quadrants of tomogram 9 of Fig.~\ref{Fig6} as the bright spot. Finally, in tomogram 10 we see how the bright region again moves toward the top quadrants, which reflects the amplification of the tidal shock arm located closer to the donor-star.

The relative brightness of the shock waves seen in the tomograms also reflects their physical difference. The hot line and the bow-shock in the disk are indeed much stronger than the arms of the tidal shock because, in the first two, the streams collide at relatively large angles, while in the tidal shock the collision is very oblique.

\subsection{Formation of the light curves}

\begin{figure}
\includegraphics[width=84mm]{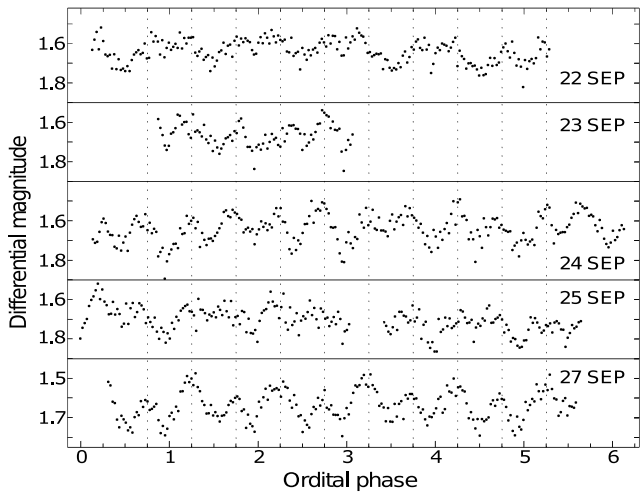}
\caption{The same light curves as in Fig.~\ref{Fig3} but the magnitude is shown as a function of orbital phase. The dashed vertical lines depict orbital phases .25 and .75.}
\label{Fig11}
\end{figure}

\rm In Fig.~\ref{Fig11} we show the individual light curves we obtained, where the magnitude is a function of the orbital phases of the binary. One can clearly see that in all five light curves the primary minima are located around phases 0.0, 1.0, etc, while the secondary minima are observed around phases 0.5, 1.5,... We observe the humps near the phases 0.25, 0.75, 1.25, 1.75, etc. We can suppose that the main contribution to the humps comes from the two strongest and, hence, brightest shock waves, directed to the observer at the phases $\phi \approx 0.2 - 0.25$ and $\phi\approx 0.75 - 0.8$, the bow-shock and the hot line respectively (designated in Fig.~\ref{Fig10} by numbers 2 and 3). However, we also see that the positions of the humps maxima vary within some range of phases near 0.25, 0.75, 1.25, 1.75, etc. from one light curve to another. The shapes and amplitudes of the humps are also different in different light curves.

\begin{figure*}
\includegraphics[width=150mm]{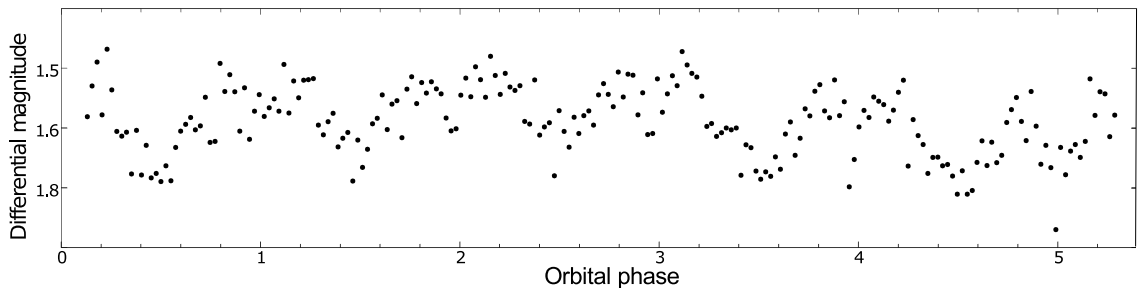}
\caption{The individual light curve of V455~And obtained on September 22, 2014.}
\label{Fig12}
\end{figure*}

Let us consider the light curve of September 22 separately shown in Fig.~\ref{Fig12}. It is interesting to note that in this light curve all the secondary minima are deeper than the primary minima, which looks unusual, since the primary minima are believed to appear because the secondary component eclipses some part of the disk and, probably, the white dwarf (the system's inclination is $i\approx75^{\circ}$ as reported by \citet{araj}). Meanwhile, in the other light curves the primary minima are deeper as it usually should be. We can explain this behavior in the following way. In Section 5.2 we have shown that, according to the analysis of the tomograms, calculated using the spectra observed on September 22 simultaneously with the photometry, the hot line is amplified when the system is observed the orbital phases 0.0, 1.0, 2.0 etc. (see also the scheme in Fig.~\ref{Fig10}), when we should see the primary minima. We suppose that the effect of shock wave amplification, resulting in somewhat increased brightness of the system, overlaps with the primary eclipse, which makes the primary minima shallower. The amplification of the bow-shock that happens, as we suppose, near the orbital phase $\phi\approx0.45$ (see Section 5.2) on September 22 is not well pronounced in the light curve, though we can see a weak feature looking like a step near the phases 0.4, 1.4, 2.4, and especially 3.4. We do not see any noticeable contribution to the light curves from the arms of the tidal shock because, as we mentioned above, these shock waves are much weaker than the hot line and the bow-shock.

It is also interesting to consider the difference between the light curve of September 22 and that of September 27. One can see that in the light curve of September 27 the humps located near orbital phases .25 are stronger than those at phases near .75. As we mentioned above, on September 22 we observe the amplification of the hot line, the strongest shock wave in the disk, near phases 1.0, 2.0, etc. as shown in the scheme in Fig.~\ref{Fig10}. The precessional density wave moves in the observer's coordinate frame with the period of the apsidal motion (see \cite{BiBo04} - \cite{Bi05}).
This period can be calculated as follows (\cite{warn}):
$$P_{am} = \frac{P_{orb}\times P_{sh}}{P_{orb} - P_{sh}}, for\ negative\ superhumps,$$
where $P_{am}$ is the period of apsidal motion, $P_{orb}$ is the orbital period, and $P_{sh}$ is the period of observed superhumps. As shown in Section 3.1 the orbital period of the system is $P_{orb} = 81.080\pm0.020$ min, and the superhump period is $P_{sh}=80.391\pm0.07$ min. This gives us the apsidal motion period $P_{am} \approx 6.57$ days. The time difference between September 22 and September 27 is 5 days or $0.76\times P_{am}$. On September 27 the density wave, in its slow retrograde (clockwise) motion, reaches the position where its outer part is directed away from the observer. This means that, when we observe the system at the phase $\phi\approx0.25$, where the bow-shock is directly exposed to us, on the opposite side of the disk the outer part of the density wave interacts with the hot line, which may additionally contribute to the hump observed at this phase. When at the phase $\phi\approx0.75$ we directly see the hot line, the bow-shock is amplified on the opposite side of the disk.

We cannot unfortunately compare the relative contribution of the amplified shock waves to the humps at the phases $\phi\approx0.25$ and $\phi\approx0.75$, since we could not obtain spectroscopic data for that night, but we can suppose that the contribution of the hot line is somewhat bigger because this shock is the strongest in the disk. This may make the hump, observed on September 27 at the phase $\phi\approx0.25$, higher than that observed at the phase $\phi\approx0.75$. 

Summarizing, we can propose the following mechanism of light curve formation. The light curves of V455 Andromedae consist of two main components. One of them is determined by the viewing aspects of the two strongest shocks in the disk, the hot line, a shock wave appearing because of the interaction of the circum-disk halo and the gas stream from the inner Lagrangian point, and the bow-shock caused by the supersonic motion of the accretor and disk in the gas of the circum-binary envelope. These shock waves produce two main humps near the orbital phases $\phi\approx0.25$ and $\phi\approx0.75$. The brightness variations of this component are modulated with the orbital period. The second component of the light curve contains up to four superhumps, occurring due to the interaction of the spiral density wave with four shocks in the disk, with the most powerful humps coming again from the hot line and bow-shock. The superhump period in our observations is $P_{sh}=80.391\pm0.069$ minutes, which is a little shorter than the orbital period. This means that in the next orbital cycle of the system each shock wave encounters the outer part of the density wave a little earlier, which makes the superhumps shift to earlier orbital phases of the system. The overlapping of two light curve components results in the shift of the humps' maxima and sometimes, as mentioned in the Introduction, may even change the number of observed humps.

\section{CONCLUSIONS}

Recently \cite{ko2015} proposed a model, based on the results of 3D gas dynamic simulations of the V455 Andromedae system, that can explain almost all the properties of WZ Sge stars' light curves. The model predicts the periodic amplification of four major shock waves in the accretion disk (hot line, two arms of the tidal shock, and bow-shock ) by a specific internal precessional density wave. This periodic amplification can cause up to four humps in the light curve of the system.

To examine the model, we conducted simultaneous spectroscopic and photometric observations of the cataclysmic variable star V455 Andromedae (WZ Sge subclass). Within 6 days, from September 22 to September 27, 2014, at the Kourovka Observatory of the Ural Federal University (Russia), we obtained 5 light curves of the system (except the night of September 26 when the weather conditions were bad). The spectroscopic observations, conducted at the Terskol Observatory (Terskol branch of The Institute of Astronomy of the Russian Academy of Sciences), resulted in obtaining 60 profiles of the $H_{\alpha}$ line on September 22, 2014. The other nights, unfortunately, were unsuccessful for spectroscopy because of poor weather conditions.

We used the $H_{\alpha}$ line profiles for Doppler tomography of the V455 And system to calculate 10 Doppler tomograms of the system that demonstrate its 10 subsequent states within one orbital period. Besides, using the results of 3D gas-dynamical modeling  of V455 And, described by \cite{ko2015}, we simulated line profiles and calculated 10 corresponding synthetic Doppler tomograms. The profiles were simulated under the assumption that the disk is optically thin. By analyzing the simulated and observed data we have found the following properties of the quiescent accretion disk in V455 And as they were on September 22, 2014:

\begin{itemize}
\item  In the simulated disk four major shock waves occur. These are: two arms of the tidal shock wave; the hot line, a shock wave occurring due to the interaction of the circum-disk halo and the gas stream from the inner Lagrangian point; and the bow-shock  caused by the super-sonic motion of the accretor and disk in the gas of the circum-binary envelope. We can also observe a one-armed spiral density wave, starting in the vicinity of the accretor and propagating to the outer regions of the disk.

\item Under the tidal action of the secondary component the density wave retrogradely precesses with an angular velocity that differs from the orbital angular velocity of the system by only 1-2\%. In course of disk's rotation each of the four shocks passes through the outer part of the density wave. This results in the periodical local density increase in the region of each shock wave. Since the energy, released in shock waves as radiation, depends on the kinetic energy of gas $\rho V^{2}/2$, by increasing the density $\rho$ we increase the energy release in the shock wave, which may be observed as increased brightness or a hump in the light curve.

\item Observed and simulated trailed spectra demonstrate similar structure and behavior. This enables us to suppose that the shape of the observed line profiles is determined by the gas-dynamical processes that we consider in our model.

\item In both the synthetic and observational Doppler tomograms we see the periodical brightening of four regions. The regions correspond to the four major shock waves in the disk, which is in a good agreement with the proposed model.

\item In the light curve of September 22 we observe that primary minima are shallower than the secondary minima. This may be explained by the fact that the hot line was amplified by the density wave when the system was observed at the orbital phase $\phi=0.0$. The resulting additional brightness overlaps with the primary eclipse, which makes the primary minimum shallower.

\item Analysis of the light curves shows that the light curves of V455 Andromedae consist of two main components. One component is determined by the viewing aspects of the two strongest shocks in the disk, the hot line, and the bow-shock. The system's brightness variations in this component are modulated with the orbital period. The second component of the light curve contains up to four superhumps occurring because of the interaction of the spiral density wave with four shocks in the disk, with the most powerful superhumps produced by the amplification of the hot line and bow-shock. These superhumps shift over the orbital phases with the period of apsidal motion of the spiral density wave. The overlapping of the main humps and the superhumps results in the shift of the humps' maxima and sometimes, may even change the number of observed humps.

\end{itemize}

The obtained observational results allow us to suppose that our model of light curve formation is correct at least for V455 And. It can explain the varying number of the humps in the light curves of WZ Sge stars, since it implies the amplification of four shock waves, potentially giving four superhumps. This effect cannot be explained in the frame of models proposed by, e.g. \citet{avil}, or \citet{wolf, Silv12}, since they imply only two sources contributing to the light curves (two-armed tidal shock by \citet{avil}, or two hot spots by \citet{wolf, Silv12}).

Our model can also explain the shift of the hump maxima, observed in different nights, by the difference between the precession angular velocity and orbital angular velocity of the system. At the next orbital cycle, the retrogradely precessing density wave interacts with the certain phase-locked shock wave a little earlier, which causes the corresponding superhump to occur at an earlier orbital phase. This may give negative superhumps. The mentioned alternative interpretations can explain neither positive nor negative superhumps since the sources of radiation they consider are phase-locked (tidal shock or hot spot). Thus, we should observe their contribution to the light curves always at the same orbital phases, which would eliminate periodicities, except the orbital variations.

The results, discussed in this paper, should, in a sense, be considered as preliminary, though giving much new information on the flow structure in WZ Sge-type stars. Further, we plan on obtaining better observational material. In particular, we, of course, need more successful nights of simultaneous spectroscopic and photometric observations to perform a more detailed analysis. Also, we need to test our model with the other representatives of WZ Sge subclass. In particular, we need to clarify the question about the positive superhumps, if they are observed in quiescent light curves of WZ Sge-type stars, since within the proposed model, implying the retrograde disk precession, we can obtain only negative superhumps.

The archive, containing animated .gif files, demonstrating the evolution of the simulated disk, "dynamic" synthetic and observational tomograms of V455 Andromedae, is available via DOI 10.13140/RG.2.1.1107.8884 on the ResearchGate (www.reaserchgate.net) web page.

\section*{Acknowledgments}
Dmitry Kononov personally thanks the Fulbright Visiting Scholar Program supported by the Department of State of the USA.
This work was partly supported by the Ministry of Education and Science of Russian Federation (the unique identifier of the project is RFMEFI59114X0003) and the Grant of the Russian President for Young Scientists (MK--2432.2013.2).

\end{document}